\documentclass[12pt]{article}
\usepackage{amsfonts,amsmath,amstext,amsthm}
\usepackage[includeheadfoot,left=1.in,right=1.in,top=1.in,bottom=1.in,bindingoffset=0.in,nohead]{geometry}
\usepackage{graphicx}
\usepackage[onehalfspacing]{setspace}
\usepackage{booktabs}
\usepackage{natbib}
\usepackage{relsize} 
\usepackage{appendix}
\usepackage[hyphens]{url}
\usepackage[pdfborder={0 0 0}]{hyperref}
\usepackage{float}
\usepackage{epstopdf}   
\usepackage{enumerate}
\usepackage{multicol} 
\usepackage{chngpage} 
 
\newcommand{\nck}[2]{\binom{#1}{#2}}

\theoremstyle{definition}
\newtheorem{remark}{Remark}
\newtheorem{theorem}{Theorem}
\newtheorem{corollary}{Corollary}
\newtheorem{algorithm}{Algorithm}

\begin{document}
\newgeometry{includeheadfoot,left=0.8in,right=0.8in,top=0.8in,bottom=0.8in,bindingoffset=0.in,nohead}

\title{On the Wisdom of Crowds \\(of Economists)} 

\author{Francis X. Diebold\\University of Pennsylvania \\and NBER \and   Aar\'{o}n Mora \\University of South Carolina \and Minchul Shin\\
Federal Reserve Bank of Philadelphia \\ $~$}

\maketitle 

\thispagestyle{empty}

\vspace{-0.5cm}
\begin{spacing}{1}

\footnotesize

\noindent \textbf{Abstract}: We study the properties of macroeconomic survey forecast response averages as the number of survey respondents grows. Such averages are ``portfolios" of forecasts. We characterize the speed and pattern of the gains from diversification as a function of portfolio size (the number of survey respondents) in both (1) the key real-world data-based environment of the U.S. Survey of Professional Forecasters, and (2) the theoretical model-based environment of equicorrelated forecast errors. We proceed by proposing and comparing various direct and model-based ``crowd size signature plots,'' which summarize the forecasting performance of $k$-average forecasts as a function of $k$, where $k$ is the number of forecasts in the average. We then estimate the equicorrelation model for growth and inflation forecast errors by choosing model parameters to minimize the divergence between direct and model-based signature plots. The results indicate near-perfect equicorrelation model fit for both growth and inflation, which we explicate by showing analytically that, under very weak conditions, the direct and fitted equicorrelation model-based signature plots are identical at a particular model parameter configuration. That parameter configuration immediately suggests an analytic closed-form estimator for the direct signature plot, so that  equicorrelation  ultimately emerges as a device for convenient calculation of direct signature plots, rather than a separate ``model" producing separate signature plots. Finally, we find that the gains from survey  diversification are greater for inflation forecasts than for growth forecasts, and that they are largely exhausted with inclusion of 5-10 representative forecasters.

\bigskip

\noindent \textbf{Acknowledgments}: For helpful comments we thank Blake LeBaron and seminar/conference participants at the University of Washington; Cornell University; the NBER Summer Institute (Cambridge, Mass.); the Conference on Real-Time Data Analysis, Methods and Applications (Bank of Canada, Ottawa); the Midwest Econometrics Group Conference (University of Kentucky); and the Conference on Macroeconomic Analysis and International Finance (University of Crete). For research assistance we thank Jacob Broussard. Remaining errors are ours alone. 

\bigskip

\noindent \textbf{Disclaimer:}: The views expressed here are those of the authors and do not necessarily represent those of the Federal Reserve Bank of Philadelphia or the Federal Reserve System.

\bigskip


\smallskip


\smallskip

\noindent {\bf Contact}:  fdiebold@sas.upenn.edu, aaron.mora@moore.sc.edu,  minchul.shin@phil.frb.org

\end{spacing}
  
  \normalsize
\restoregeometry
  
\newpage
\thispagestyle{empty}
\setcounter{tocdepth}{3} 
\tableofcontents

\newpage
\setcounter{page}{1}
\thispagestyle{empty}

\section{Introduction and Basic Framework} \label{intro}
 
The wisdom of crowds, or lack thereof, is traditionally and presently a central issue in  psychology, history, and political science; see for example \cite{surowiecki2005}  regarding wisdom, and \cite{kindleberger2023} regarding lack thereof. Perhaps most prominently, however, the wisdom of crowds is also---and again, traditionally and presently---a central issue in economics and finance, where heterogeneous information and expectations formation play a prominent role.\footnote{Interestingly, moreover, it also features prominently in new disciplines like machine learning and artificial intelligence, via forecast combination methods like ensemble averaging \citep[e.g.,][]{DSZ2023}.}

In this paper we focus on economics and finance, studying the ``wisdom" of ``crowds" of professional economists. We focus on the U.S. Survey of Professional Forecasters (SPF), which is important not only in facilitating empirical academic research in macroeconomics and financial economics, but also---and crucially---in  guiding real-time policy, business, and investment management decisions.\footnote{On real-time policy and its evaluation, see for example John Tayor's inaugural NBER Feldstein Lecture at \url{https://www.hoover.org/sites/default/files/gmwg-empirically-evaluating-economic-policy-in-real-time.pdf}.}$^,$\footnote{For an introduction to the SPF, see the materials at 
        \url{https://www.philadelphiafed.org/surveys-and-data/real-time-data-research/survey-of-professional-forecasters}.} 

In particular, we study SPF crowd behavior as crowd size grows, asking precisely the same sorts of questions of SPF  ``forecast portfolios"  that one asks of financial asset portfolios:\footnote{Classic early work on which we build includes, for example, \cite{makridakis1983averages} and \cite{BD1995}. We progress much further, however, particularly regarding analytic characterization.} How quickly, and with what patterns, do  diversification benefits become operative, and eventually dissipate, as portfolio size (the number of forecasters) grows, and why? Do the results differ across the key SPF variables (growth and inflation), and if so, why?  What are the implications for survey size and design?

We answer the above questions using what we call ``crowd size signature plots," which summarize the forecasting performance of $k$-average forecasts as a function of $k$, where $k$ is the number of forecasts in the average. We examine both ``direct signature plots" (estimated directly from the SPF data) and ``model-based signature plots" (estimated by fitting a simple model of forecast-error equicorrelation), and we characterize their surprisingly intimate relationship, uncovering en route a remarkably simple analytic closed-form signature plot estimator. We focus throughout on estimating and interpreting signature plots for both growth and inflation, providing individual and comparative assessments of their paths and patterns.

\paragraph{Related Literature.}


This paper contributes to the forecast combination literature by examining the performance patterns of simple $k$-average forecasts as the number of forecasters, $k$, increases. We note that this paper is not focused on the problem of optimal forecast combination (see \cite{BatesandGranger1969} for a pioneering contribution, and for example \cite{timmermann2006forecast}, \cite{hsiao2014} and \cite*{wang2022} for recent developments). 
The motivation for focusing on simple $k$-averages is twofold.
First, it is a well-established empirical fact that simple averages of forecasts produce combined forecasts with surprisingly good forecasting performance despite not being optimal in general (e.g., \cite{clemen1989},  \cite{stock2004combination}, and \cite*{genre2013}).\footnote{This empirical fact has been called a ``forecast combination puzzle''; work focused on studying the causes of this puzzle includes \cite{smith2009simple}, \cite{Elliott2025}, and \cite*{claeskens2016simple}.}
Second, simple averages can be optimal in specific contexts. For example, \citet{EngleandKelly2012} show that when forecast errors exhibit equicorrelation, equal weights are optimal; \citet{Elliott2025} later provide precise necessary and sufficient conditions under which simple averages are optimal.



Our focus on performance patterns of simple k-averages as k increases builds on earlier empirical work. \citet{makridakis1983averages}, using the M-Competition time-series dataset and combinations of forecasts across multiple forecasting methods, document that simple-average combinations become more accurate, with sharply diminishing marginal gains, as the number of component forecasts increases. We find the same qualitative pattern in the SPF when averaging across professional forecasters. We extend these results by providing a theoretical foundation: we derive an exact expression for the direct crowd size signature plot that links its level and slope to two interpretable features of SPF forecast errors (average dispersion and average dependence). This delivers an equivalent equicorrelation representation and explains both the hyperbolic shape of the signature plots and the rapid exhaustion of diversification gains.

In another related paper, \cite{chan2018some} propose a framework to study the properties of a combined forecast, and derive conditions under which the simple average is optimal. In their setting, they point out that the performance of a combined forecast improves as the number of forecasts or models increases, and that a simple average will asymptotically outperform any fixed individual forecast.
Although this is consistent with our results, our emphasis is more general, as we characterize the entire trajectory of performance improvements, including the practically relevant setting of a small to moderate number of forecasters.
 
Our results on relating direct crowd size signature plots to an equicorrelation model-based signature plot are connected to the results in \cite{clemen1985limits}. In the case of normally distributed forecast errors, they show that the value of additional information for forecast combination diminishes when the additional information is correlated. This is consistent with our findings documenting diversification gains; however, our results go beyond this by not requiring the normality assumption.


Finally, we contribute to the SPF survey-forecast literature (e.g., \citealp{CroushoreStark2019}; \citealp{clements2022}) by providing evidence and theory directly relevant for survey design: how many forecasters are needed to capture most of the benefits of averaging, and why do those benefits differ across variables? Using SPF growth and inflation forecasts, we document a steeply diminishing-returns pattern in crowd-size performance, with most improvements realized by about 5-10 representative forecasters. Moreover, we show that diversification gains are considerably larger for inflation than for growth, and we provide an exact theoretical characterization linking the inflation-growth difference to the underlying variance-covariance structure of individual forecast errors.

We proceed as follows. In section \ref{databased} we study the SPF, estimating its crowd size signature plots directly, for both growth and inflation.  In section \ref{model} we introduce and estimate the equicorrelation model, beginning in section \ref{model1} by characterizing its crowd size signature plots analytically for any parameter configuration, and continuing in section \ref{estimate} by estimating its parameters by minimizing divergence between direct and model-based signature plots, again providing empirical results for both growth and inflation. In section \ref{concl} we conclude and sketch several directions for future research.

\section{Direct SPF Crowd Size  Signature Plots} \label{databased}

\begin{figure}[tbp]
	\caption{SPF Participation}
	\label{fig: n of resp} 
	\begin{center}
		\includegraphics[trim = 00mm 00mm 0mm 00mm,clip,scale=.65]{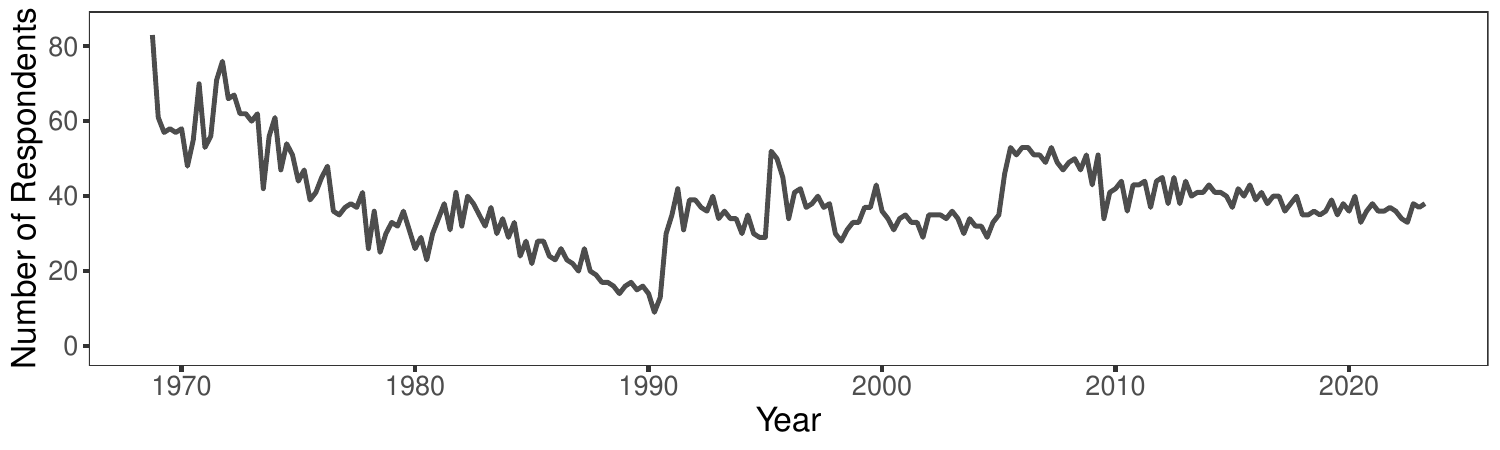}
	\end{center}
	\vspace{-3mm}
		\begin{spacing}{1.0}  \noindent  \footnotesize 
			Notes: We show the number of participants in the U.S. Survey of Professional Forecasters, 1968Q4-2023Q2. 
		\end{spacing}
\end{figure}

In this section we introduce the idea of crowd size signature plots, and we calculate them directly for SPF point forecasts of real output growth (``growth") and GDP deflator inflation (``inflation"), for  forecast horizons $h=1,2, 3, {\rm and ~} 4$, corresponding to short-, medium-, and longer-term forecasts.\footnote{The SPF contains quarterly level forecasts of real GDP and the GDP implicit price deflator.  We transform the level forecasts into growth and inflation  forecasts by computing annualized quarter-on-quarter growth rates, and we compute the corresponding forecast errors using realized values as of  December 2023. See Appendix \ref{app_data} for details.}  Our sample period is  1968Q4-2023Q2.

The SPF was started in 1968Q4 and is currently  conducted and maintained by the Federal Reserve Bank of Philadelphia.\footnote{For a recent introduction see  \cite{CroushoreStark2019}.} In Figure \ref{fig: n of resp} we show the evolution of the number of forecast participants, which declined until 1990Q2, when the Federal Reserve Bank of Philadelphia took control of the survey, after which it has had approximately 40 participants.  Participants stayed for 15 quarters on average, with a minimum of 1 quarter and a maximum of 125 quarters.


We us now sketch the basic framework. Let  $N$ refer to a   set of forecasts  with $N \times 1$ zero-mean time-$t$ error vector $e_t$, $t=1, ..., T$, and let $k \le N$ refer to a subset of  forecasts.  We consider $k$-forecast averages, and we seek to characterize $k$-forecast  mean-squared forecast error ($MSE$). For a particular  $k$-forecast average corresponding to group $g^*_k$, the forecast error is just the average of the individual forecast errors, so we have 
\begin{equation}
	\widehat{MSE}^*_T (k) = { \frac{1}{T}\sum_{t=1}^{T}   \left(\frac{1}{k}  \sum_{i \in g^*_k}  e_{it} \right)^{2}}. \label{mse}
\end{equation}
For any choice of $k$, however, there are $N \choose k$ possible $k$-forecast   averages. We  focus on the $k$-average $\widehat{MSE}^*_T(k)$ given by equation \eqref{mse}, averaged across all groups of size $k$,  
\begin{equation}
	\widehat{MSE}^{avg}_{NT}(k) = \frac{1}{{N \choose k}}  \sum_{g_k =1}^{{N \choose k}} \left ( \frac{1}{T} \sum_{t=1}^{T}  \left(\frac{1}{k} \sum_{i \in g_k} e_{it} \right)^{2} \right ), \label{MSE2}
\end{equation}
where $g_k$ is an arbitrary member  of the set of groups of size $k$.

Among other things, one may be interested in:

\begin{enumerate}
	
	\item \label{top} Tracking and visualizing $\widehat{MSE}^{avg}_{NT}(k)$ as $k$ grows (``$\widehat{MSE}^{avg}_{NT}(k)$ crowd size   signature plots");

	\item Tracking and visualizing \textit{functions} of  $\widehat{MSE}^{avg}_{NT}(k)$ as $k$ grows, such as: 
	
	\begin{enumerate}

		\item the  performance from $k$-averaging ($\widehat{MSE}^{avg}_{NT}(k)$) relative to the performance from no averaging ($\widehat{MSE}^{avg}_{NT}(1)$),  
		\begin{equation}  \label{key1}
			\widehat{MSE}^{avg}_{R,NT}(k) = \frac{\widehat{MSE}^{avg}_{NT}(k)}{\widehat{MSE}^{avg}_{NT}(1)}
		\end{equation}
		(``$\widehat{MSE}^{avg}_{R,NT}(k)$ crowd size   signature plots," where ``$R$" denotes ``relative," or ``ratio");
		
		\item the \textit{change} (improvement) in  $ \widehat{MSE}^{avg}_{NT}(k)$ from adding one more forecast to the  pool (i.e., moving from $k$ to $k+1$ forecasts), e.g.,
		\begin{equation}
			\widehat{DMSE}^{avg}_{NT}(k)  = \widehat{MSE}^{avg}_{NT}(k)  - \widehat{MSE}^{avg}_{NT}(k+1)  \label{dmse}
		\end{equation}
		(``$\widehat{DMSE}^{avg}_{NT}(k)$ crowd size signature plots," where ``$D$" denotes ``delta," or ``change"),
		
		or, in ratio form,
\begin{equation}
	\widehat{DMSE}^{avg}_{R,NT}(k) = \frac{\widehat{DMSE}^{avg}_{NT}(k)}{\widehat{DMSE}^{avg}_{NT}(1)}
\end{equation}
		(``$\widehat{DMSE}^{avg}_{R,NT}(k)$ crowd size  signature plots");
				
	\end{enumerate}

	\item Tracking and visualizing not just the  \textit{mean} of the squared-error distribution  as $k$ grows, as in all of the above signature plots, but rather the \textit{complete}  distribution (``$\widehat{F}^{avg}_{NT}(k)$  signature plots", or in ratio form, ``$\widehat{F}^{avg}_{R,NT}(k)$ signature plots");
	
	\item Estimating  equicorrelation models for SPF forecast errors by minimizing divergence between direct and model-based SPF signature plots, and assessing their fit;
	
	\item Examining the paths and  patterns in the above estimated signature plots, and similarities/differences for growth vs. inflation;
	
	\item Drawing practical implications for SPF design.
	
\end{enumerate} 

\noindent The ``ratio" signature plots, especially the $\widehat{MSE}^{avg}_{R,NT}(k)$ plot defined in equation \eqref{key1}, deserve special mention.  Scaling by $\widehat{MSE}^{avg}_{NT}(1)$ (the benchmark \textit{MSE} corresponding to no averaging) makes  $\widehat{MSE}^{avg}_{R,NT}(1) \equiv 1$, which facilitates $\widehat{MSE}^{avg}_{NT}(k)$ comparisons  across different variables like growth and inflation. For that reason we use it extensively in the graphical presentations that follow.

$\widehat{MSE}^{avg}_{NT}(k)$ and the related objects above are readily calculated in principle, but complications  arise in practice.  In particular,  calculation is infeasible unless  $N$ is very small, due to the huge number of different $k$-average forecasts. For example, for $N=40$ and $k=20$ we obtain ${N \choose k} =   O(10^{11})$.   Hence we proceed by approximating  $\widehat{MSE}^{avg}_{NT}(k)$ as follows:

\begin{algorithm} \quad  \label{algsim}
	
\begin{enumerate}
	
	   \item \textit{For each period $t=1,...,T$, randomly select $k$ forecasts and average their forecast errors; refer to this $k$-average forecast group as $g^*_{k,t}$;}

       \item \textit{Average the forecast errors for $g^*_{k,t}$ across sample periods to  calculate $\widehat{MSE}^*_T (k)$;}
          
        \item \textit{Repeat  $B$ times, and average the $\widehat{MSE}^*_T (k)$ values across the $B$ draws, where $B$ is large, but not so large as to be computationally intractable.}
\end{enumerate} 
\end{algorithm}

\noindent  As $B \rightarrow \infty$, the above approximation will converge to $\widehat{MSE}^{avg}_{NT}(k)$, and in this paper we set $B=30,000$.  The approximation works, moreover, for unbalanced as well as balanced panels, which is important because the SPF panel is unbalanced due to entry and exit of forecasters. 

\begin{figure}[tbp]
	\caption{Direct $\widehat{MSE}^{avg}_{R,NT}(k)$ Crowd Size Signature Plots}
	\label{fig: RMSE}
	\begin{center} 
		\textbf{Growth}         ~~~~~~~~~~~~~~~~~~~~~~~~~~~~~~~~~~~~~~~~~~~~~~~~                        \textbf{Inflation}
		\includegraphics[trim = 00mm 2mm 0mm 0mm,clip,scale=.6]{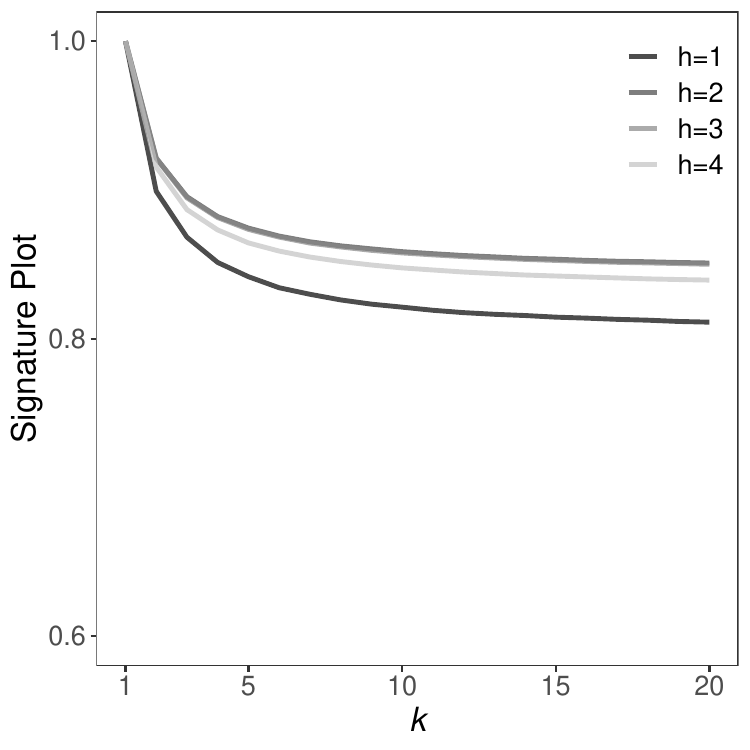}
		\includegraphics[trim = 00mm 2mm 0mm 0mm,clip,scale=.6]{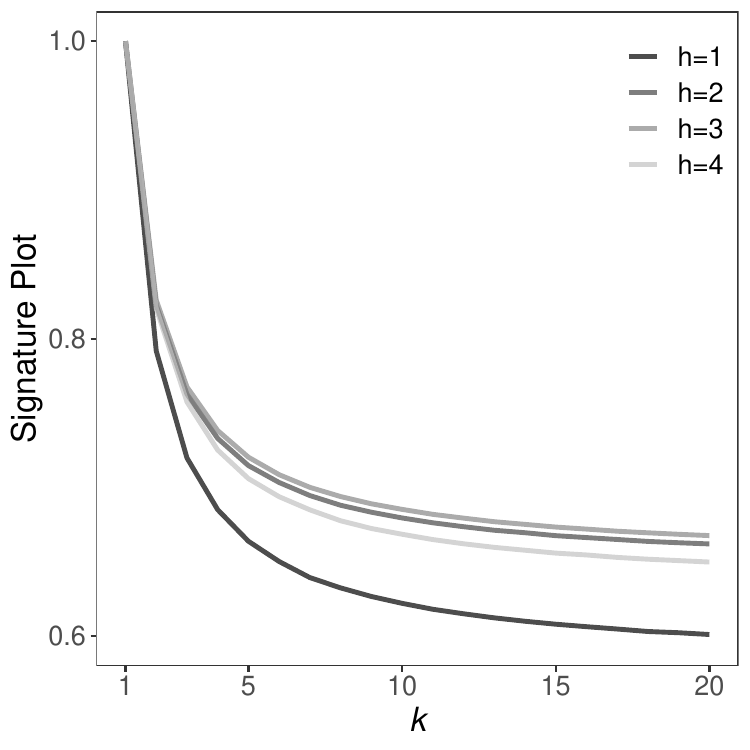}
	\end{center}
	\vspace{-2mm}
	\begin{spacing}{1.0}  \noindent  \footnotesize 
		Notes: We show direct $\widehat{MSE}^{avg}_{R,NT}(k)$ crowd size signature plots for SPF growth and inflation forecasts at  horizon $h=1,2,3,4$, for group sizes $k=1, 2, ..., 20$. For each $k$, we produce the figure by randomly drawing $B=30,000$ groups of size $k$ for each $t=1,...,T$.
	\end{spacing}   
\end{figure}

\begin{figure}[tbp]
	\caption{Direct $\widehat{DMSE}^{avg}_{R,NT}(k)$  Crowd Size Signature Plots  }
	\label{fig: DMSE}
	\begin{center}
		\textbf{Growth}         ~~~~~~~~~~~~~~~~~~~~~~~~~~~~~~~~~~~~~~~~~~~~~~~~                \textbf{Inflation}
		\includegraphics[trim = 00mm 2mm 0mm 0mm,clip,scale=0.6]{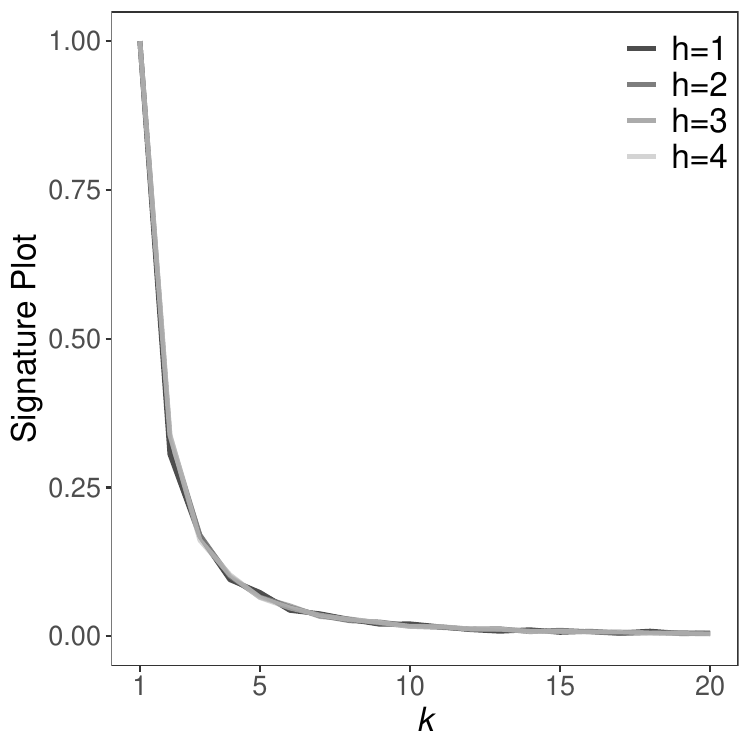}
		\includegraphics[trim = 00mm 2mm 0mm 0mm,clip,scale=0.6]{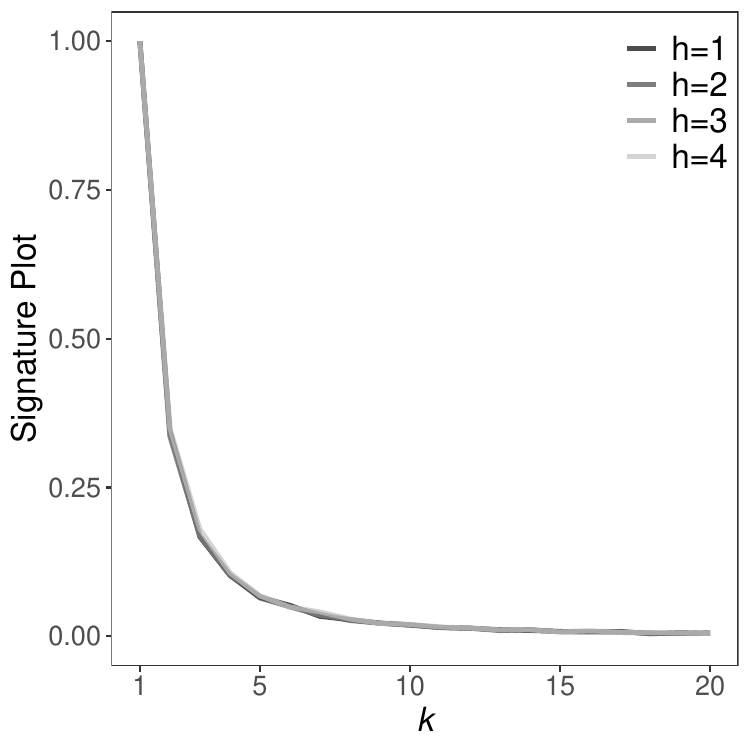}
	\end{center}
	
	\vspace{-2mm}
	\begin{spacing}{1.0}  \noindent  \footnotesize 
		Notes:  We show direct $\widehat{DMSE}^{avg}_{R,NT}(k)$ crowd size  signature plots for SPF growth and inflation forecasts at  horizons $h=1,2,3,4$, for group sizes $k=1, 2, ..., 20$. For each $k$, we produce the figure by randomly drawing $B=30,000$ groups of size $k$ for each $t=1,...,T$.
	\end{spacing}   
\end{figure}

\begin{figure}[t]
	\caption{Direct $\widehat{F}^{avg}_{R,NT}(k)$ Crowd Size Signature Plots}
	\label{fig: boxplots}
	\begin{center}
		
		\textbf{Growth}         ~~~~~~~~~~~~~~~~~~~~~~~~~~~~~~~~~~~~~~~~~~~~~~~~                        \textbf{Inflation}
		
		%
		%
		
		\includegraphics[trim = 00mm 2mm 0mm 0mm,clip,scale=.63]{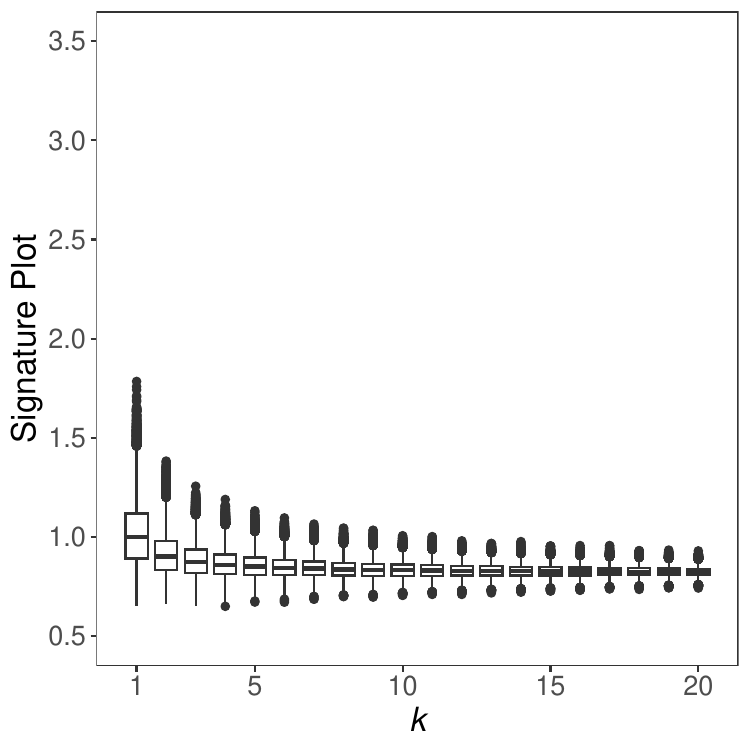}
		\includegraphics[trim = 00mm 2mm 0mm 0mm,clip,scale=.63]{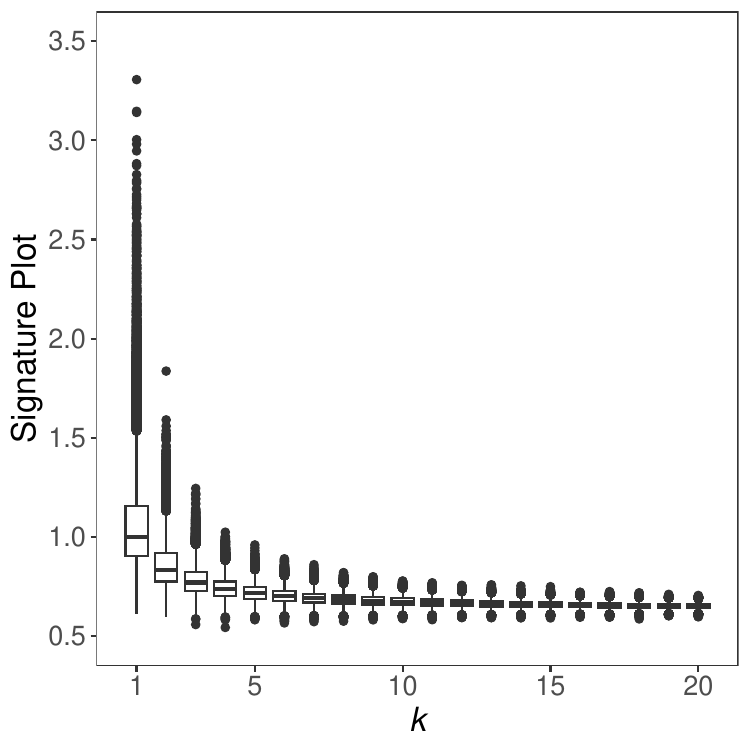}
	\end{center}
	
	\vspace{-2mm}
	\begin{spacing}{1.0}  \noindent  \footnotesize 
		Notes: We show direct $\widehat{F}^{avg}_{R,NT}(k)$ crowd size signature plots for SPF growth and inflation forecasts.  The signature plots are composed of boxplots of squared 1-step-ahead $k$-average forecast errors, $k=1, 2, ..., 20$. Each boxplot displays the median, the first and third quartiles, the lower extreme value (first quartile minus 1.5 times interquartile range), the upper extreme value (third quartile plus 1.5 times the interquartile range), and outliers.  All growth boxplots and all inflation boxplots are scaled by their respective medians at $k=1$, so that both the growth and inflation medians equal 1.0 when $k=1$. For each $k$, we produce the figure by randomly drawing $B=30,000$ groups of size $k$ for each $t=1,...,T$.
	\end{spacing}   
\end{figure}

We show  direct  crowd size signature plots for growth and inflation in Figure \ref{fig: RMSE} ($\widehat{MSE}^{avg}_{R,NT}(k)$), Figure \ref{fig: DMSE} ($\widehat{DMSE}^{avg}_{R,NT}(k)$), and Figure \ref{fig: boxplots} ($\widehat{F}^{avg}_{R,NT}(k)$). Several features are apparent, as distilled in the following remarks:

\begin{remark}
For both growth and inflation, the $\widehat{MSE}^{avg}_{R,NT}(k)$ signature plot is lowest for $h=1$, with the signature plots for $h$ = 2, 3, and 4  progressively farther above the $h=1$ plot in roughly parallel upward shifts, reflecting the fact that the near future is generally easier to predict than the more-distant future.
\end{remark}

\begin{remark}
The growth $\widehat{MSE}^{avg}_{R,NT}(k)$ signature plot does not rise much from $h=3$ to $h=4$, in contrast to the inflation $\widehat{MSE}^{avg}_{R,NT}(k)$ signature plot,  suggesting that growth predictability drops with horizon more quickly than inflation predictability, effectively vanishing by $h=3$.
\end{remark}

\begin{remark}
For both growth and inflation and all forecast horizons, the reduction in $\widehat{MSE}^{avg}_{R,NT}(k)$ from $k=1$ to $k=5$ dwarfs the improvement from moving from $k=6$ to $k=20$, as visually emphasized by the $\widehat{DMSE}^{avg}_{R,NT}(k)$ signature plots in Figure \ref{fig: DMSE}.  Hence there is little benefit from adding representative forecasters to the pool beyond $k=5$.
\end{remark}

\begin{remark}
 Individual examination of the growth or inflation  $\widehat{DMSE}^{avg}_{R,NT}(k)$ signature plots reveals approximately identical behavior at all horizons (i.e., no upward shifts), which is expected because the $\widehat{MSE}^{avg}_{R,NT}(k)$ signature plots shift with horizon in approximately parallel fashion, leaving its ``first derivative" plot  ($\widehat{DMSE}^{avg}_{R,NT}(k)$) approximately unchanged.
\end{remark}

\begin{remark}
Comparison  of the growth and inflation $\widehat{DMSE}^{avg}_{R,NT}(k)$ signature plots reveals  identical marginal gains from adding additional representative forecasters. This phenomenon is not accidental, and we shall return to it subsequently in section \ref{model}. 
\end{remark}

\begin{remark}  \label{remf}
	Both the growth and inflation $\widehat{F}^{avg}_{R,NT}(k)$ plots show right-skewness in forecast-error distributions for small \textit{k}, but the distributions become less variable and more symmetric as \textit{k} grows and the law of large numbers and central limit theorem operate. The small-$k$ skewness is much greater, however, for the inflation forecast-error distribution. 
\end{remark}

\begin{remark} \label{rem}
The growth vs. inflation $\widehat{MSE}^{avg}_{R,NT}(k)$ plots asymptote to very different levels as  $k$ increases -- approximately 80\% for growth and 60\% for inflation -- which highlights a central result: The benefits of SPF ``portfolio diversification" appear to be substantially greater for inflation than for growth.
\end{remark}

Figures \ref{fig: RMSE} and \ref{fig: boxplots}, together with Remarks \ref{remf} and \ref{rem}, are closely linked and merit additional and joint discussion. Figure \ref{fig: boxplots} and Remark \ref{remf} show a markedly stronger consensus in growth forecasts than in inflation forecasts (compare, for example, the growth and inflation forecast distributions in Figure \ref{fig: boxplots} for $k=1$). This pattern accords with the broader empirical macroeconomics literature, in which low-ordered autoregressions typically provide hard-to-beat  descriptions of short-term U.S. growth \citep[e.g.,][]{SW2002,CP2013}, whereas inflation dynamics are more complex, warranting consideration of a wider set of models \citep[e.g.,][]{SW2007,PR2007}.\footnote{A similar situation appears in the agent-based modeling literature, where simulated agents struggle to learn a stable inflation model and often adopt heterogeneous specifications \citep[e.g.,][]{ABK2013,LS2025}.} In this light, Figure \ref{fig: RMSE} and Remark \ref{rem} make perfect sense: the greater heterogeneity in inflation forecasts produces greater gains from inflation forecast diversification, yielding a strikingly lower $\widehat{MSE}^{avg}_{R,NT}(k)$ signature-plot asymptote for inflation (about 60\%) than for growth (about 80\%).

\section{Model-Based SPF Crowd Size Signature Plots} \label{model}

Having empirically characterized crowd size signature plots directly in the SPF data, we now proceed to characterize them analytically in a simple covariance-stationary equicorrelation model, in which $e_{t} \sim (0, \Sigma)$, 
where $0$ is the $N \times 1$ zero vector and $\Sigma$ is an  $N\times N$  forecast-error covariance matrix displaying equicorrelation, by which we mean that all variances are identical and all correlations are identical. 

\subsection{Population Results Under Equicorrelation} \label{model1}

A trivial equicorrelation example occurs when $\Sigma = \sigma^2 I$, where $I$ denotes the $N\times N$ identity matrix, so that all variances are equal ($\sigma^2$), and all correlations are equal (0). Of course the zero-correlation case is unrealistic, because, for example, 
 economic forecast errors are invariably positively correlated due to overlap of information sets, but it will serve as a useful benchmark, so we begin with it. 

Simple averaging is the fully optimal forecast combination in the zero-correlation environment, which is obvious since the forecasts are exchangeable. More formally, the optimality of simple averaging (equal combining weights) follows from the multivariate \cite{BatesandGranger1969} formula for $MSE$-optimal combining weights, 
\begin{equation}  \label{optimal1}
        \lambda^* = \left (\iota' \Sigma^{-1} \iota \right)^{-1}  \Sigma^{-1} \iota,
\end{equation}
where $\iota$ is a $k$-dimensional column vector of ones. For $\Sigma = \sigma^{2} I$ the optimal weights collapse to
$$
\lambda^* = (\sigma^{-2} N)^{-1} \sigma^{-2} \iota= \frac{1}{N} \iota. \nonumber 
$$

Analytical asymptotic results are straightforward for simple averages in the zero-correlation environment. Let $MSE^{avg}_{N \infty}(k;   \sigma) =   plim_{T \rightarrow \infty}  \left (  \widehat{MSE}^{avg}_{NT}(k; \sigma) \right )$; then we have 
\begin{equation}  \label{eqn4}
        MSE^{avg}_{N \infty}(k;   \sigma) =  plim_{T \rightarrow \infty} \left ( \frac{1}{{N \choose k}}  \sum_{g_k =1}^{{N \choose k}} \left ( \frac{1}{T} \sum_{t=1}^{T}  \left(\frac{1}{k} \sum_{i \in g_k} e_{it} \right)^{2} \right ) \right )
\end{equation}
\begin{equation*}
= \frac{1}{{N \choose k}}  \sum_{g_k =1}^{{N \choose k}} \left ( E  \left[ \left(\frac{1}{k} \sum_{i \in g_k} e_{it} \right)^{2} \right] \right)
\end{equation*}
\begin{equation*} 
=\frac{1}{{N \choose k}}  \sum_{g_k =1}^{{N \choose k}} \frac{1}{k^2}  E  \left[ \left( \sum_{i \in g_k} e_{it} \right)^{2} \right] 
\end{equation*}
\begin{equation*}
        =\frac{1}{{N \choose k}}  \sum_{g_k =1}^{{N \choose k}} \frac{1}{k^2} \sum_{i \in g_k} E\left[  e_{it}^{2} \right] 
\end{equation*}
\begin{equation*}
        = \frac{\sigma^{2}}{k}. 
\end{equation*}

\noindent Moreover,
\begin{equation}
        DMSE^{avg}_{N \infty}(k; \sigma) = \frac{ \sigma^2 }{k(k+1)}
\end{equation}
and 
\begin{equation} \label{eqn44}
        MSE^{avg}_{R,N \infty}(k) = \frac{1}{k}.
\end{equation}
Notice that $\sigma$ cancels in the $R^{avg}_{N \infty}(k; \sigma)$ calculation, so we simply write $R^{avg}_{N \infty}(k)$.

We now move to a richer equicorrelation case with equal but nonzero correlations, so that instead of $\Sigma = \sigma^{2} I$ we have
\begin{equation}
\Sigma = \sigma^{2} R, \label{equi0a}
\end{equation}
where
\begin{equation} \label{equi0}
R = 
\begin{pmatrix}
        1 & \rho & \cdots & \rho \\
        \rho & 1 & \cdots & \rho \\
        \vdots & \vdots & \ddots & \vdots \\
        \rho & \rho & \cdots & 1 \\ 
\end{pmatrix},
\end{equation}  
and $\rho \in \left ] \frac{-1}{N-1}, 1 \right [$.\footnote{$R$ is positive definite if and only if $\rho \in \left ] \frac{-1}{N-1}, 1 \right [$.  See Lemma 2.1 of \cite{EngleandKelly2012}.}  Recent work, in particular \cite{EngleandKelly2012}, has made use of equicorrelation in the context of modeling multivariate financial asset return volatility.

Importantly, the optimality of simple averaging under zero correlation is preserved under equicorrelation.  That is, equicorrelation is sufficient for the optimality of simple averaging -- an immediate implication of the results of \cite{Elliott2025}, who shows that a necessary and sufficient condition for optimality of simple averaging is that row sums of $\Sigma$ be equal.  Equicorrelation is one such case, although there are of course others, obtained by manipulating  correlations in their relation to variances to keep row sums equal, but none are nearly so compelling and readily interpretable as equicorrelation.

To understand the optimality of simple averaging under equicorrelation in the context of \cite{BatesandGranger1969}, just as we did earlier under zero correlation, consider the inverse covariance matrix in the Bates-Granger expression for the optimal combining weight vector, \eqref{optimal1}.  In  the equicorrelation case we have
\begin{equation} \label{equi_matrix}
        \Sigma^{-1} = \frac{1}{\sigma^{2}}  R^{-1},
\end{equation}
where\footnote{See Lemma 2.1 of \cite{EngleandKelly2012}.}
\begin{equation} \label{equi_matrix2}
        R^{-1} = \frac{1}{1-\rho} I - \frac{\rho}{(1-\rho)(1+(N-1)\rho)} \iota \iota'.
\end{equation}
 Then, using equation (\ref{equi_matrix2}), the first part of the optimal combining weight \eqref{optimal1} is
\begin{equation} \label{eq: eq proof 1}
        \iota' \Sigma^{-1} \iota  = \frac{N}{\sigma^{2}} \frac{(1+(N-1)\rho)- \rho N}{(1-\rho)(1+(N-1)\rho)},
\end{equation}
and the second part is
\begin{equation} \label{eq: eq proof 2}
        \Sigma^{-1} \iota = \frac{1}{\sigma^{2}} \frac{(1+(N-1)\rho) - \rho N}{ (1-\rho)(1+(N-1)\rho)} \iota .
\end{equation}
Inserting  equations   (\ref{eq: eq proof 1}) and (\ref{eq: eq proof 2}) into  equation  \eqref{optimal1} again yields
\begin{equation}
        \lambda^* = \frac{1}{N} \iota,
\end{equation}
establishing the optimality of equal weights.

Having now introduced equicorrelation and shown that it implies optimality of simple average forecast combinations, it is of interest to motivate it -- despite its stark simplicity -- in terms of economic considerations.  First, obviously but importantly, the information sets of economic forecasters are quite highly overlapping, so it is not necessarily unreasonable to suppose that the forecast error variances are roughly equal, and that pairs of forecast errors are roughly equally correlated.

Second, less obviously but very importantly,  equicorrelation is closely linked to factor structure, which is a great workhorse of modern macroeconomic and business-cycle analysis \citep[e.g.,][]{SW2}. In particular, equicorrelation arises when forecast errors have single-factor structure with equal factor loadings and equal idiosyncratic shock variances, as for example in:
\begin{equation}
        e_{it} = \delta z_{t} + w_{it} \label{facmod}
\end{equation}
\vspace{-10mm}
$$
z_{t} = \phi z_{t-1} + v_{t},
$$
where $ w_{it} \sim iid (0, \sigma_{w}^{2})$, $v_{t} \sim iid (0, \sigma_{v}^{2})$, and $w_{it} \perp v_{t}$,  $\forall i, t$, $i = 1, ..., N$, $t = 1, ..., T$.

Finally, a large literature from the 1980s onward documents and exploits the routine outstanding empirical performance of simple average forecast combinations, despite the fact that simple averages are not optimal in general \citep[e.g., ][]{clemen1989, genre2013, ET2016, DS2019}. As we have seen, however,  equicorrelation is sufficient (and almost necessary) for optimality of simple averages, so that if simple averages routinely perform well, then the equicorrelation model is routinely reasonable -- and a natural model to pair with the simple averages embodied in the SPF.

Analytical results for $MSE^{avg}_{N \infty}(\cdot)$, $DMSE^{avg}_{N \infty}(\cdot)$ and $R^{avg}_{N \infty}(\cdot)$ are easy to obtain under equicorrelation, just as they were under zero correlation.  Immediately,
\begin{equation}
        \label{MSEequi}
        MSE^{avg}_{N \infty}(k;  \rho, \sigma) = plim_{T \rightarrow \infty} \left ( \frac{1}{{N \choose k}}  \sum_{g_k =1}^{{N \choose k}} \left ( \frac{1}{T} \sum_{t=1}^{T}  \left(\frac{1}{k} \sum_{i \in g_k} e_{it} \right)^{2} \right ) \right )
\end{equation}
        \begin{equation*}
= \frac{1}{{N \choose k}}  \sum_{g_k =1}^{{N \choose k}} \left ( E  \left[ \left(\frac{1}{k} \sum_{i \in g_k} e_{it} \right)^{2} \right] \right)
\end{equation*}
\begin{equation*}
        =\frac{1}{{N \choose k}}  \sum_{g_k =1}^{{N \choose k}} \frac{1}{k^2}  E  \left[ \left( \sum_{i \in g_k} e_{it} \right)^{2} \right] 
\end{equation*}
\begin{equation*}
        =\frac{1}{{N \choose k}}  \sum_{g_k =1}^{{N \choose k}} \frac{1}{k^2} \left[ k\sigma^2+ k(k-1) Cov(e_{it},e_{jt}) \right] \quad \text{for } \: i\neq j
\end{equation*}
\begin{equation*}
     =\frac{\sigma^{2}}{k} \left[ 1+(k-1) \rho \right]. 
\end{equation*}
In addition, for $MSE$ changes and ratios we have
\begin{equation} \label{eq: value of k (rmse)}
        DMSE^{avg}_{N \infty}(k;  \rho,  \sigma) =  \frac{\sigma^2}{k(k+1)}(1-\rho)
\end{equation}
and
\begin{equation} \label{rrrrr}
        MSE^{avg}_{R,N \infty}(k; \rho, \sigma) = MSE^{avg}_{R,N \infty}(k; \rho) = \frac{1}{k} \left[ 1+(k-1) \rho \right].
\end{equation}
One can immediately verify that if $\rho=0$ then the equicorrelation signature plot results \eqref{MSEequi}-\eqref{rrrrr} collapse to the corresponding earlier zero-correlation results \eqref{eqn4}-\eqref{eqn44}.

\begin{figure}[t]
        \caption{Theoretical Equicorrelation $MSE^{avg}_{R,N \infty}(k; \rho)$ Crowd Size Signature Plots}
        \label{fig: equicorr R and M}
        \begin{center}
                \includegraphics[trim = 00mm 00mm 0mm 00mm,clip,scale=.45]{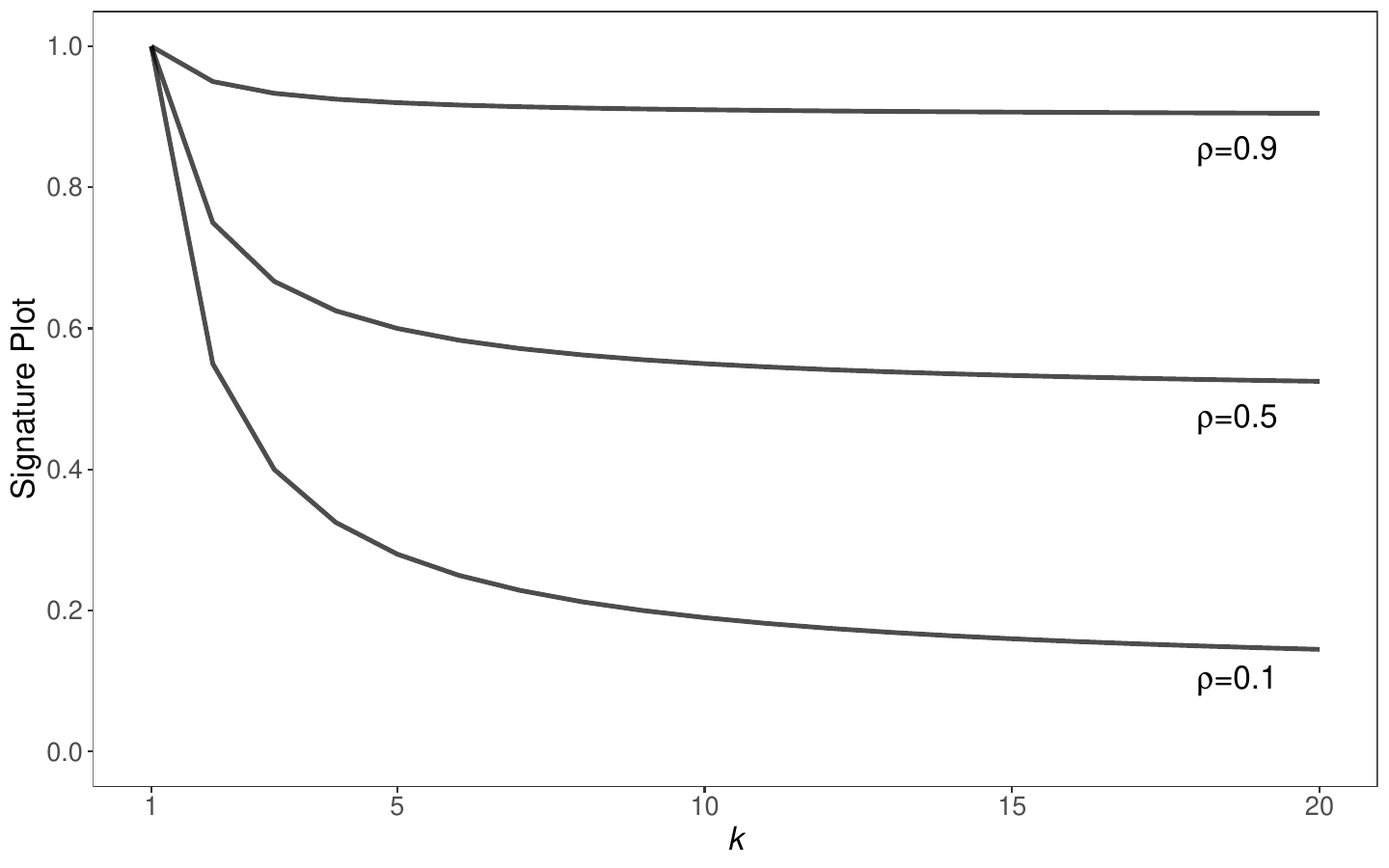}
        \end{center}
        \begin{adjustwidth}{1cm}{1cm}
\vspace{-2mm}
                \begin{spacing}{1.0}  \noindent  \footnotesize 
                        Notes: We show theoretical equicorrelation $MSE^{avg}_{R,N \infty}(k; \rho)$ crowd size signature plots for  equicorrelations $\rho = 0.1, 0.5, 0.9$ and group sizes $k = 1, ..., 20$.        
                \end{spacing}
        \end{adjustwidth}
        \vspace{1mm}
\end{figure}

\begin{figure}[t]
        \caption{Theoretical Equicorrelation $MSE^{min}_{R,N \infty}(k;  .5)$, $MSE^{avg}_{R,N \infty}(k;   .5)$, and $    MSE^{max}_{R,N \infty}(k;    .5)$ Crowd Size Signature Plots}
        \label{fig:sim_MSE}
        \begin{center}
                \includegraphics[trim = 00mm 0mm 0mm 00mm,clip,scale=.45]{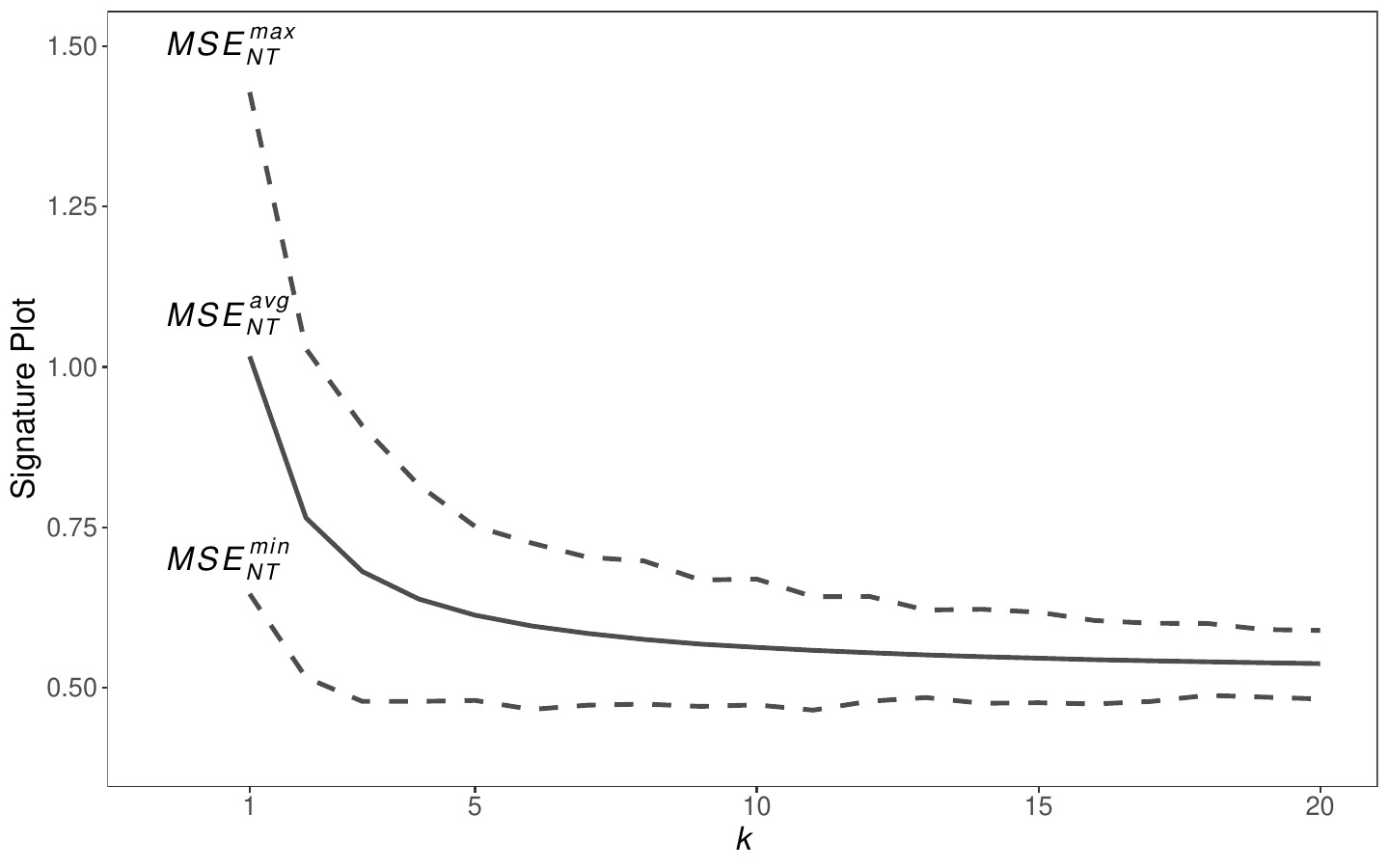}
        \end{center}    
        \begin{adjustwidth}{1cm}{1cm}
                \vspace{-2mm}
                \begin{spacing}{1.0}  \noindent  \footnotesize 
                        Notes: The data generating process is the equicorrelation model given by equations \eqref{equi0a}-\eqref{equi0}, with $\rho=.5$, $\sigma=1$, $N=40$, and $T=160$.  For each group size $k$, 30,000 groups of size $k$ were drawn at random for each $t=1,...,T$ to produce the figure.
                \end{spacing}   
        \end{adjustwidth}
        \vspace{1mm}
\end{figure}

In Figure \ref{fig: equicorr R and M} we show  $MSE^{avg}_{R,N \infty}(k;  \rho)$ as a function of $k$, for various equicorrelations, $\rho$. Several remarks are in order: 

\begin{remark}
	From equation \eqref{rrrrr}, the $MSE^{avg}_{R,N \infty}(k;  \rho)$ curves  decrease  to a limiting value ($\rho$) as the combining pool grows ($k \rightarrow \infty$). Indeed the gains from increasing $k$ are initially large but decrease hyperbolically quickly. The $MSE$ improvement, for example, in moving from $k=1$ to $k=5$ consistently dwarfs that of moving from $k=5$ to $k=20$.  
\end{remark}

\begin{remark}
Overall, the value of averaging across forecasters is determined by the interaction of $k$ and $\rho$. Other things equal, it is highest when $k$ is small (small pool) or when $\rho$ is low (weakly correlated forecast errors).  In particular, for realistic values of $\rho$, around 0.5, say, most gains from increasing $k$ are obtained by $k=5$.  
\end{remark}

\begin{remark}
 The fact that, for realistic values of $\rho$, most $MSE^{avg}_{R,N \infty}(k; \rho)$  gains from increasing $k$ are obtained by $k=5$ does \textit{not} necessarily indicate  that typical surveys use too many forecasters. $MSE^{avg}_{R,N \infty}(k; \rho)$ is an \textit{average} across all $k$-forecast combinations, and the best and worst $k$-average combinations, for example, will have very different \textit{MSE}s. 
Figure \ref{fig:sim_MSE} speaks to this; it shows $MSE^{min}_{R,N \infty}(k;  \rho)$, $MSE^{avg}_{R,N \infty}(k;  \rho)$, and $     MSE^{max}_{R,N \infty}(k;  \rho)$) under equicorrelation with $\rho=0.5$, for $k=1, ..., 20$.\footnote{We use $N=40$ as an approximation to the average number of forecasters participating in the SPF in any given quarter, and we use $T=160$ to mimic the total sample size when working with 40 years of quarterly data, as in the SPF.}
\end{remark}

\begin{remark}
The equicorrelation case is the only one for which analytic results are readily obtainable.  For example, even if we maintain the assumption of equal correlations but simply allow different forecast-error variances (``weak equicorrelation"), the $MSE$ of the $k$-person average forecast becomes a function of $(k, \rho, \sigma_{1}^{2}, ..., \sigma_{N}^{2})$, and little more can be said.\footnote{See Appendix \ref{genequi} for derivation of optimal combining weights in the weak equicorrelation case.}  
\end{remark}

\begin{remark} 
	As mentioned earlier,  the equicorrelation case naturally matches the provision of survey averages, because in that case simple averages are optimal. Hence, as we now proceed to a model-based empirical analysis of real forecasters, we work with the equicorrelation model, asking what values of $\rho$ and $\sigma$  make the equicorrelation model-based $MSE^{avg}_{N \infty}(k;  \rho, \sigma)$ signature plot as close as possible to the direct $\widehat{MSE}^{avg}_{NT}(k)$ signature plot.
\end{remark}

\subsection{Estimating the Equicorrelation Model} \label{estimate}

Here we estimate the equicorrelation model by choosing its parameters $\rho$ and $\sigma$ to make the equicorrelation $MSE^{avg}_{N \infty}(k;  \rho, \sigma)$  as close as possible to the SPF $\widehat{MSE}^{avg}_{NT}(k)$. This estimation strategy is closely related to, but different from, GMM estimation. Rather than matching model and data moments, it matches more interesting and interpretable functions of those moments, namely model-based and direct crowd size signature plots  -- as per the ``indirect inference" of \cite{Smith1993} and \cite{Gourierouxetal1993}. Henceforth we refer to it simply as the ``matching estimator."

Specifically, we solve for  $(\hat{\rho}, \hat{\sigma})$ such that
\begin{equation}   \label{obj_fn}
  (\hat{\rho},\hat{\sigma})=\arg \min_{(\rho, \sigma)}  Q(\rho, \sigma),
\end{equation}
where 
\begin{equation*}
        Q(\rho, \sigma)=\frac{1}{N} \sum_{k=1}^N \left(\widehat{MSE}^{avg}_{NT}(k)- MSE_{N \infty}^{avg}(k; \rho, \sigma) \right)^2 ,
\end{equation*}
and the minimization is constrained such that $\sigma>0$ and $\rho \in \left] \frac{-1}{N-1},1 \right [$.

\subsubsection{Calculating the Solution}

\begin{figure}[tp]
        \caption{Illustration of the Relationship Between $\rho$ and $\sigma^2$}
        \label{fig:match_est_growth2}
        \begin{center}
                \includegraphics[trim = 00mm 3mm 0mm 0mm,clip,scale=.45]{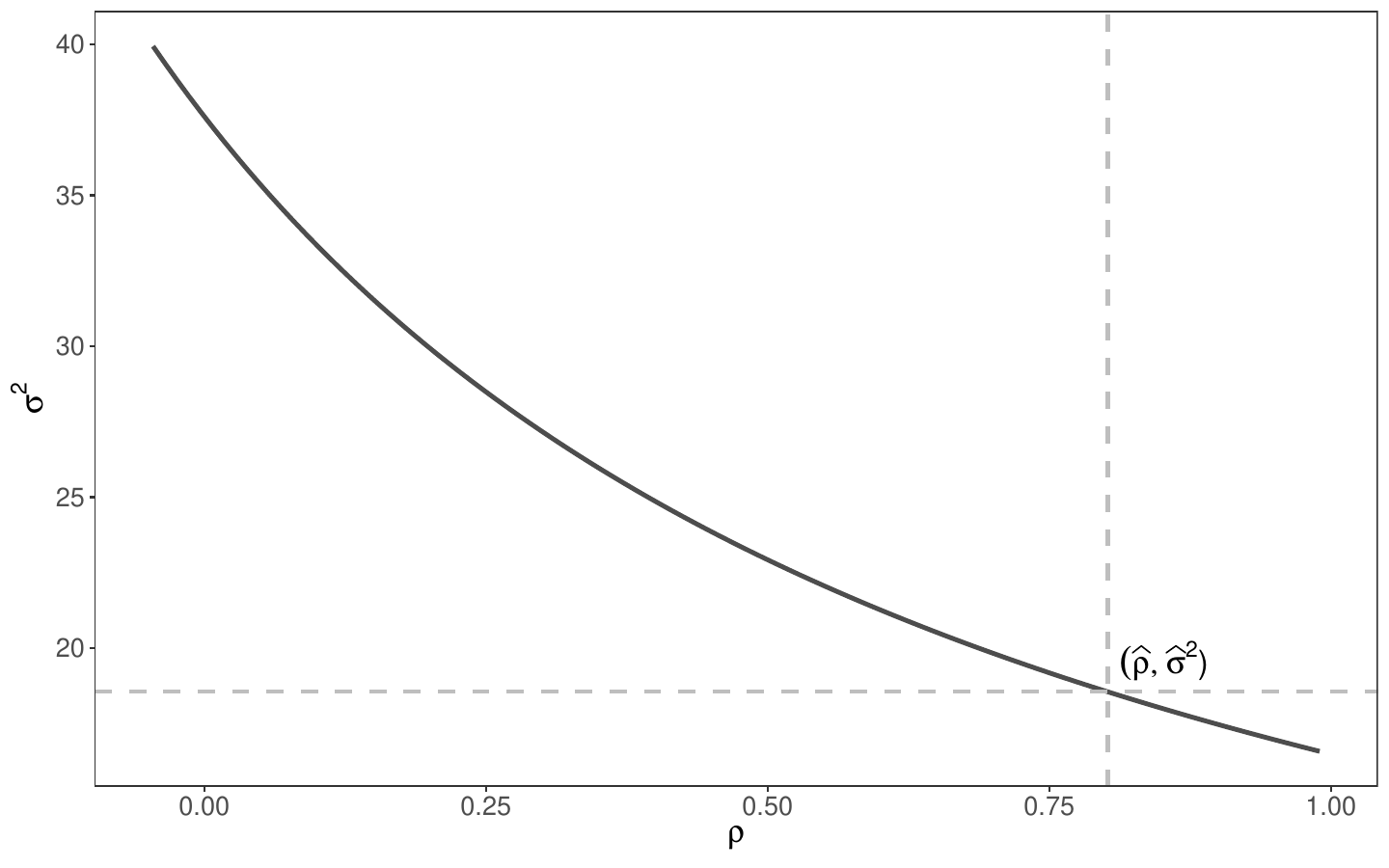}               
        \end{center}
        \begin{adjustwidth}{1cm}{1cm}
                \vspace{-2mm}
                \begin{spacing}{1.0}  \noindent  \footnotesize 
                        Notes: We show the relationship  between  $\rho$  and $\sigma^2$ given by equation \eqref{rel_sigma_rho}, for 1-step-ahead growth forecast errors. The highlighted values of $\hat{\rho}=0.8$ and $\hat{\sigma}^{2}=18.6$ are the estimated values for growth forecasts at horizon $h=1$. See Table \ref{tbl_match_est_output_growth}. 
                \end{spacing}   
        \end{adjustwidth}
\end{figure}

\begin{figure}[tp]
        \caption{Illustration of the Objective Function and its Minimum}
        \label{fig:match_est_growth}
        \begin{center}
                \begin{tabular}{cc}
                        \quad \quad  $Q(\rho)$ &  \quad \quad \quad  $Q(\rho)$ in a Neighborhood of $\hat{\rho}$ \\
                        \includegraphics[trim = 00mm 3mm 0mm 0mm,clip,scale=0.55]{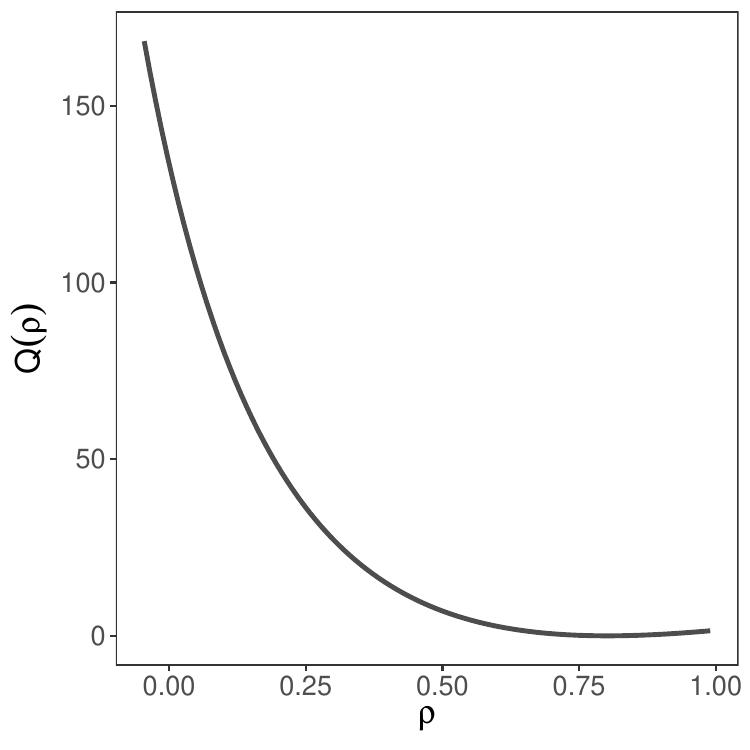} & 
                        \includegraphics[trim = 00mm 3mm 0mm 0mm,clip,scale=0.55]{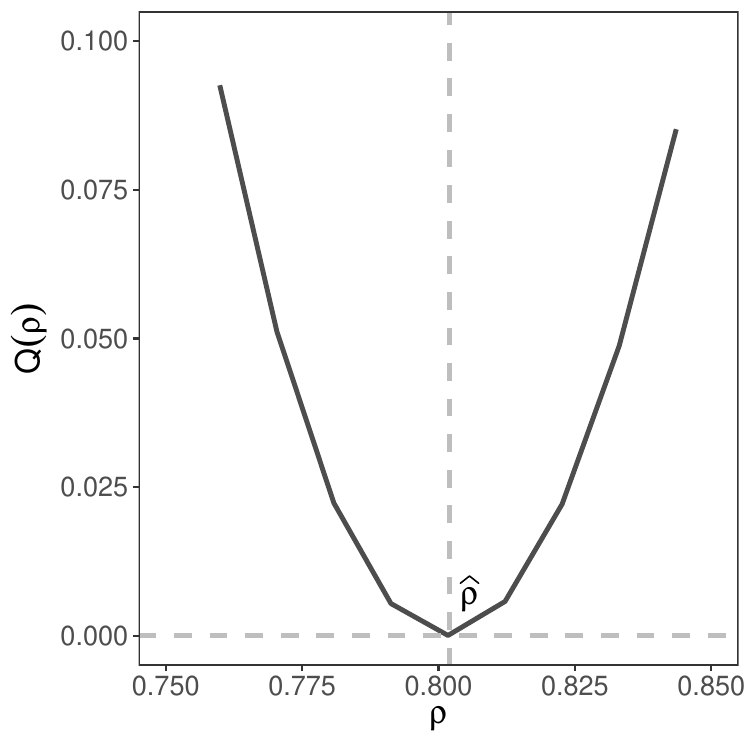}              
                \end{tabular}
        \end{center}
                \begin{adjustwidth}{1cm}{1cm}
                        \vspace{-2mm}
                \begin{spacing}{1.0}  \noindent  \footnotesize 
                Notes: We show the objective function $Q(\rho)$ of the matching estimator expressed as a function of $\rho$, for 1-step-ahead growth forecast errors. The highlighted value of $\hat{\rho}=0.8$ is the estimated value for growth forecasts at horizon $h=1$. See Table \ref{tbl_match_est_output_growth}.                
                \end{spacing}
        \end{adjustwidth}       
\end{figure}

Calculating the minimum in equation \eqref{obj_fn} is very simple, because the bivariate minimization can be reduced to a univariate minimization in $\rho \in \left] \frac{-1}{N-1},1 \right [$. To see this, recall that $MSE_{N \infty}^{avg}(k; \rho, \sigma)=\frac{\sigma^2}{k}[1+(k-1)\rho]$, so that the first-order condition for $\sigma^2$ is
\begin{equation} \label{eq1}
 \frac{1}{N} \sum_{k=1}^N \left(\widehat{MSE}^{avg}_{NT}(k)- MSE_{N \infty}^{avg}(k; \rho, \sigma) \right)\left( \frac{1}{k}+ \frac{k-1}{k} \rho\right) =0,
\end{equation}
and the first-order condition for $\rho$ is
\begin{equation} \label{eq2}
 \frac{1}{N} \sum_{k=1}^N \left(\widehat{MSE}^{avg}_{NT}(k)- MSE_{N \infty}^{avg}(k; \rho, \sigma) \right)\left( 1-\frac{1}{k} \right)\sigma^2 =0.
\end{equation}  
Combining equations \eqref{eq1} and \eqref{eq2} yields
\begin{equation}
        \label{rel_sigma_rho}
        \sigma^2=\frac{c_1}{c_2+c_3\rho},  
\end{equation}
where $c_1=\sum_{k=1}^N \frac{\widehat{MSE}^{avg}_{NT}(k)}{k}$, $c_2=\sum_{k=1}^N \frac{1}{k^2}$ and $c_3=\sum_{k=1}^N \frac{k-1}{k^2}$. Hence, at the optimum and conditional on the data, there is a deterministic inverse relationship between $\rho$ and $\sigma$ (see Figure \ref{fig:match_est_growth2}), enabling one to restrict the parameter search to the small open interval $\rho \in \left] \frac{-1}{N-1},1 \right [$, as well as to explore the objective function visually as a function of $\rho$ alone (see Figure \ref{fig:match_est_growth}).

\begin{table}[tp]
        \caption{Equicorrelation Model Estimates}
        \label{tbl_match_est_output_growth}
        \begin{center}
                \begin{tabular}{rcccc}
                        \multicolumn{5}{c}{\textbf{Growth}}   \\
                        \hline \hline
                        \\[-1.8ex] & $h=1$    & $h=2$    & $h=3$    & $h=4$    \\
                        \hline
                        \\[-1.8ex] $\hat{\sigma}^2$                                & 18.562   & 21.170   & 22.713   & 23.275   \\
                        & (0.606)    & (0.546)    & (0.589)    & (0.646)    \\
                        $\hat{\rho}$                                    & 0.801    & 0.843    & 0.842    & 0.831    \\
                        & (0.036)    & (0.028)    & (0.028)    & (0.030)    \\
                        ${MSE}^{avg}_{R,N \infty}(1;   \hat{\rho})$  & 1.000    & 1.000    & 1.000    & 1.000    \\
                        ${MSE}^{avg}_{R,N \infty}(5;   \hat{\rho})$  & 0.841    & 0.874    & 0.874    & 0.865    \\
                        ${MSE}^{avg}_{R,N \infty}(15;  \hat{\rho})$ & 0.815    & 0.853    & 0.853    & 0.842    \\
                        &          &          &          &          \\
                        $Q(\hat{\sigma}^2,\hat{\rho})$                  & 5.99E-05 & 3.95E-06 & 7.95E-06 & 1.01E-05 \\
                        \hline
                        &          &          &          &          \\
                        \multicolumn{5}{c}{\textbf{Inflation}}                                                           \\
                        \hline
                        \\[-1.8ex]& $h=1$    & $h=2$    & $h=3$    & $h=4$    \\
                        \hline
                        \\[-1.8ex] $\hat{\sigma}^2$                                & 3.662    & 4.343    & 5.123    & 6.094    \\
                        & (0.253)    & (0.255)    & (0.295)    & (0.370)    \\
                        $\hat{\rho}$                                    & 0.580    & 0.644    & 0.650    & 0.630    \\
                        & (0.082)    & (0.068)    & (0.066)    & (0.071)    \\
                        ${MSE}^{avg}_{R,N \infty}(1;   \hat{\rho})$  & 1.000    & 1.000    & 1.000    & 1.000    \\
                        ${MSE}^{avg}_{R,N \infty}(5;   \hat{\rho})$  & 0.664    & 0.715    & 0.720    & 0.704    \\
                        ${MSE}^{avg}_{R,N \infty}(15;  \hat{\rho})$ & 0.608    & 0.668    & 0.673    & 0.655    \\
                        &          &          &          &          \\
                        $Q(\hat{\sigma}^2,\hat{\rho})$                  & 2.21E-06 & 9.56E-07 & 1.82E-06 & 4.36E-05 \\
                        \hline\hline
                \end{tabular}
        \end{center}
        \begin{adjustwidth}{1cm}{1cm}
        \begin{spacing}{1.0}  \noindent  \footnotesize 
                Notes: We show equicorrelation model parameter estimates for SPF growth and inflation forecast errors at various horizons, with standard errors computed via 1000 bootstrap samples. We also show estimated relative $MSE$ with respect to no averaging, ${MSE}^{avg}_{R,N \infty}(k;\hat{\rho})$  for $k=1,5,15$. In the final line of each panel we show the value of the objective function evaluated at the estimated parameters, $Q(\hat{\sigma}^2,\hat{\rho})$.
        \end{spacing}
        \end{adjustwidth}
\end{table}


\begin{figure}[tp]
        \caption{Direct $\widehat{MSE}_{R,NT}^{avg}(k)$ and Estimated Equicorrelation  ${MSE}_{R,N \infty}^{avg}(k; \hat{\rho})$ Crowd Size Signature Plots}
        \label{fig: Fitted Signature Plots_111}
        \vspace{4mm}
        \vspace{-2mm} 
        \begin{center}
                \textbf{Direct, Growth}     \quad \quad   \quad \quad  \quad \quad  \quad \quad  \quad   \quad \quad  \textbf{Est. Equicorrelation, Growth}
                \includegraphics[trim={0cm 0cm 0 0},clip,scale=0.6]{graphics/Figure_9a_cs.pdf} 
                \includegraphics[trim={0cm 0cm 0 0},clip,scale=0.6]{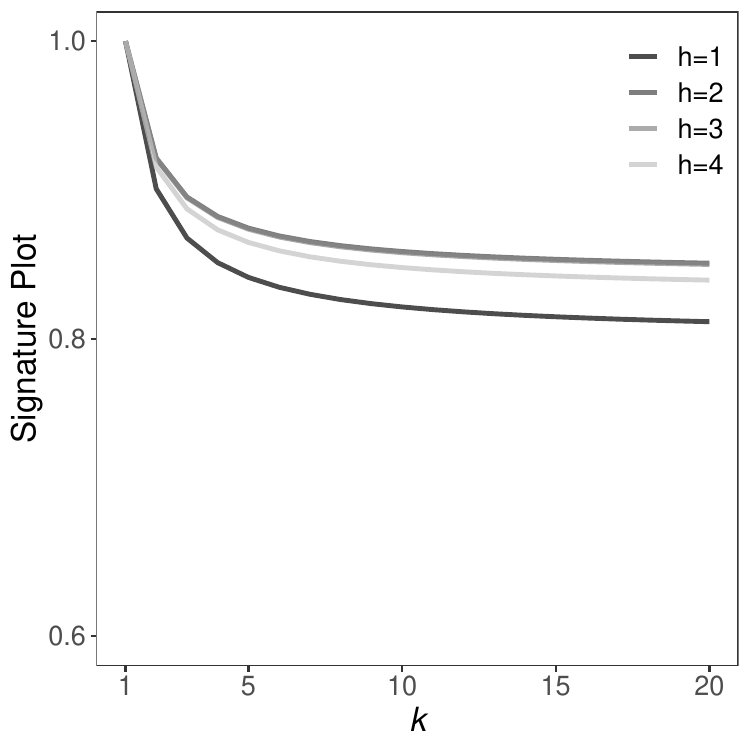}
                \textbf{Direct, Inflation}     \quad \quad   \quad \quad  \quad \quad  \quad \quad  \quad   \quad \quad  \textbf{Est. Equicorrelation, Inflation}
                \includegraphics[trim={0cm 0cm 0 0},clip,scale=0.6]{graphics/Figure_9b_cs.pdf} 
                \includegraphics[trim={0cm 0cm 0 0},clip,scale=0.6]{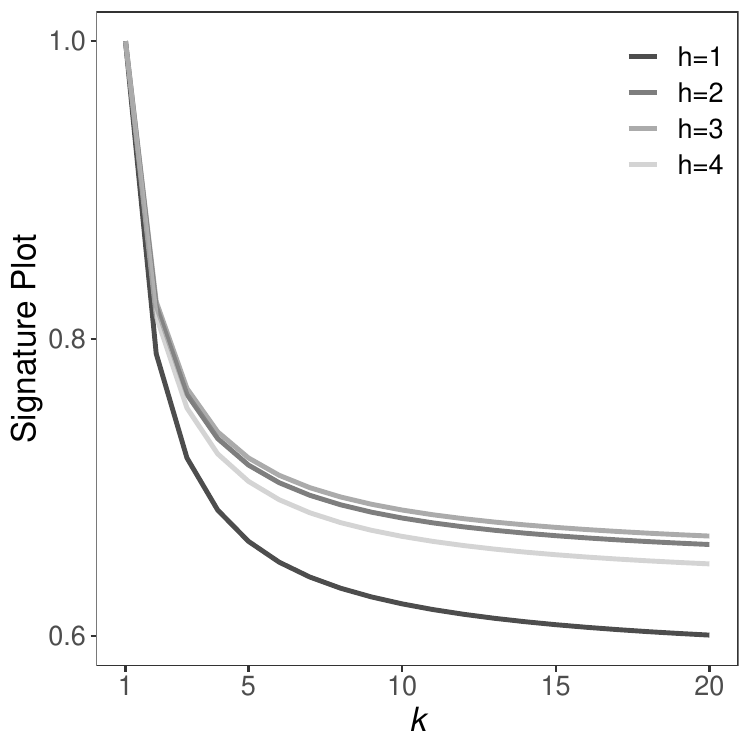}    
                
        \end{center}    
        \vspace{-2mm}
        \begin{adjustwidth}{1cm}{1cm}
                \begin{spacing}{1.0}  \noindent  \footnotesize 
                        Notes: We show direct and estimated equicorrelation ratio crowd size signature plots ($\widehat{MSE}_{R,NT}^{avg}(k)$ and ${MSE}_{R,N \infty}^{avg}(k; \hat{\rho})$, respectively)  for growth and inflation forecasts at horizons $h=1,2,3,4$, for group sizes $k=1, 2, ..., 20$.         
                \end{spacing}
        \end{adjustwidth}               
\end{figure}

In Table \ref{tbl_match_est_output_growth} we show the complete set of estimates (for ${\sigma}^2$ and ${\rho}$, for growth and inflation, for $h=1, ...4$). The $\hat{\rho}$ estimates accord with the earlier-discussed asymptotes of the direct signature plots, and the  $\hat{\sigma}^2$ estimates increase with forecast horizon, reflecting the fact that the distant future is harder to forecast than the near  future, and implying that fitted model-based equicorrelation signature plots should shift upward with horizon. In Figure \ref{fig: Fitted Signature Plots_111} we show side-by-side direct $\widehat{MSE}^{avg}_{NT}(k)$ (left column) and equicorrelation model-based ${MSE}^{avg}_{R,N \infty}(k;  \hat{\rho})$ (right column)  signature plots, which reveal a remarkably good equicorrelation model fit -- the direct and model-based plots are effectively identical.

\subsubsection{Understanding the Near-Perfect Equicorrelation Fit}

Here we present a closed-form solution for the direct crowd size signature plot. The result is significant in its own right and reveals why our numerical matching estimates for the equicorrelation model produce fitted signature plots that align so closely with direct signature plots. To maintain precision it will prove useful to state it as a formal theorem.  

\vspace{1mm}

\begin{theorem} \label{bigtheorem}

\textit{Let $e_{t}$ be any covariance stationary $N \times 1$ vector with mean zero and covariance matrix $\Sigma$, given by 
\[
\mathbf{\Sigma} =
\begin{pmatrix}
        \sigma_{1}^{2} & c_{12} & c_{13} & \cdots & c_{1N} \\
        c_{21} & \sigma_{2}^{2} & c_{23} & \cdots & c_{2N} \\
        c_{31} & c_{32} & \sigma_{3}^{2} & \cdots & c_{3N} \\
        \vdots & \vdots & \vdots & \ddots & \vdots \\
        c_{N1} & c_{N2} & c_{N3} & \cdots & \sigma_{N}^{2}
\end{pmatrix},
\]
and define the population $k$-average $MSE$,
\[
MSE^{avg}_{N \infty}(k) = \frac{1}{\nck{N}{k}} \sum_{g_{k}=1}^{\nck{N}{k}} E\left[\left(\frac{1}{k} \sum_{i \in g_{k}} e_{it}\right)^{2}\right],
\]
where $g_{k}$ represents any subset of $e_t$ of size $k$ ($k \in [1, N]$).  Then
\begin{equation} \label{mmm}
MSE_{N \infty}^{avg}(k) = \frac{\overline{\sigma^{2}}}{k}\left( 1  + (k-1) \overline{\rho} \right)
\end{equation}
and
\begin{equation} \label{rrr}
MSE^{avg}_{R,N \infty}(k) = \frac{MSE^{avg}_{N \infty}(k)}{MSE^{avg}(1)} = \frac{1}{k} \left( 1 + (k-1) \overline{\rho} \right),
\end{equation}
where
\[
\overline{\sigma^{2}} = \frac{1}{N} \sum_{i=1}^{N} \sigma_{i}^{2}
\]
\[
\overline{c}= \frac{1}{\nck{N}{2}} \sum_{1\leq i <j\leq N} c_{ij}
\]
\[
\overline{\rho} = \frac{\overline{c}}{\overline{\sigma^{2}}}.
\]
}\end{theorem}

\begin{proof}
	We have:
$$
MSE^{avg}_{N \infty}(k) = \frac{1}{\nck{N}{k}} \sum_{g_{k}=1}^{\nck{N}{k}} E\left[\left(\frac{1}{k} \sum_{i \in g_{k}} e_{it}\right)^{2}\right]
$$
$$ 
= \frac{1}{k^{2}} \frac{1}{\nck{N}{k}} \sum_{g_{k}=1}^{\nck{N}{k}} \left[ \sum_{i \in g_{k}} \sigma_{i}^{2} + \sum_{i,j \in g_{k}, i\neq j} c_{ij}\right]
$$
$$
= \frac{1}{k^{2}} \frac{1}{\nck{N}{k}} \left[ \underbrace{\sum_{g_{k}=1}^{\nck{N}{k}}  \sum_{i \in g_{k}} \sigma_{i}^{2}}_{\text{grand sum of variances}} + \underbrace{\sum_{g_{k}=1}^{\nck{N}{k}} \sum_{i,j \in g_{k}, i\neq j} c_{ij}}_{\text{grand sum of covariances}}\right].
$$
 
\noindent First consider the term related to variance. Note that, of the $\nck{N}{k}$ groups, there are $\nck{N-1}{k-1}$ groups that include $e_{i}^{2}$. The sum of $\sum_{i\in g_{k}} \sigma_{i}^{2}$ over all groups $g_{k}$ is therefore
\begin{equation} \label{term1}
\sum_{g_{k}=1}^{\nck{N}{k}} \sum_{i \in g_{k}} \sigma_{i}^{2} = \nck{N-1}{k-1} \sum_{i=1}^{N} \sigma_{i}^{2}.
\end{equation}

\noindent Now consider the term related to covariance. When summing $\sum_{i,j \in g_{k}, i\neq j} c_{ij}$ across all groups $g_{k}$, covariances between all possible pairs of $e_{it}$ and $e_{jt}$ are accounted for. Because we are summing over all arbitrary groups $g_{k}$, each pair $(i,j)$ appears the same number of times in the grand summation. To compute this number we observe that each group $g_{k}$ contains $k(k-1)$ pairwise covariances and that there are $\nck{N}{k}$ possible groups. Hence the total number of individual covariance terms in the grand sum is $k(k-1) \nck{N}{k}$. The number of times that each individual covariance term $c_{ij}$ appears in the grand sum is $\frac{k(k-1)\nck{N}{k}}{\nck{N}{2}}$, where $\nck{N}{2}$ is the total number of distinct pairs $(i,j)$. The grand sum of covariances is therefore
\begin{equation} \label{term2}
\frac{1}{\nck{N}{k}} \sum_{g_{k}=1}^{\nck{N}{k}} \sum_{i,j \in g_{k}, i\neq j} c_{ij} = \frac{k(k-1)\nck{N}{k}}{\nck{N}{2}} \sum_{i \leq i < j \leq N} c_{ij}.
\end{equation}

\noindent Combining equations \eqref{term1} and \eqref{term2}, we have:
$$
MSE^{avg}_{N \infty}(k)
= \frac{1}{k^{2}} \frac{1}{\nck{N}{k}} \nck{N-1}{k-1} \sum_{i=1}^{N} \sigma_{i}^{2} + \frac{1}{k^{2}} \frac{1}{\nck{N}{k}}   \frac{k(k-1)\nck{N}{k}}{\nck{N}{2}} \sum_{i \leq i < j \leq N} c_{ij}
$$
$$
= \frac{1}{k}\left[ \left(\frac{1}{N}\sum_{i=1}^{N} \sigma_{i}^{2} \right) +(k-1) \left( \frac{1}{\nck{N}{2}} \sum_{i \leq i < j \leq N} c_{ij} \right) \right] 
$$
$$
= \frac{\overline{\sigma^{2}}}{k} \left[1 + (k-1) \overline{\rho} \right].
$$

\noindent This completes the proof.
\end{proof}

Several remarks are in order:

\begin{remark} \label{equiv}
 Theorem \ref{bigtheorem} reveals that in large samples the direct crowd size signature plot is simply the equicorrelation model-based signature plot evaluated at particular values of the equicorrelation model parameters $\rho$ and $\sigma$, \textit{regardless of whether the forecast errors are truly equicorrelated}. Hence in large samples the two plots will always match perfectly. The same identity holds exactly in finite samples upon replacing expectations by time averages, using $\widehat{\sigma}_{i}^{2}=T^{-1}\sum_{t=1}^{T} e_{it}^{2}$ and $\widehat{c}_{ij}=T^{-1}\sum_{t=1}^{T} e_{it}e_{jt}$ when the panel is balanced.  
\end{remark}

\begin{remark}  \label{estima}
	 Theorem \ref{bigtheorem} also immediately suggests an alternative,  closed-form, matching estimator for the equicorrelation model: 
	\begin{equation}          \label{alt2}
		\widehat{\sigma}^{2} = \frac{1}{N} \sum_{i=1}^{N} \widehat{\sigma}_{i}^{2}
	\end{equation}
	\begin{equation} \label{alt1}
		\widehat{\rho} = \frac{\frac{1}{\nck{N}{2}} \sum\limits_{1\leq i <j\leq N} \widehat{c}_{ij}}{\frac{1}{N} \sum\limits_{i=1}^{N} \widehat{\sigma}_{i}^{2}}. 
	\end{equation}
\end{remark}		

\begin{remark}
Combining Remarks \ref{equiv} and \ref{estima}, it emerges that, regardless of whether one chooses to approach the signature plot from the ``direct side" or from the ``equicorrelation side," there is no need for simulation (as earlier from  the direct side, when implementing the direct estimator via Algorithm  \ref{algsim}) or numerical optimization (as earlier from the equicorrelation side, when implementing the matching estimator via solution of  the first-order conditions \eqref{eq1}-\eqref{eq2}). Instead, the trivially simple closed-form estimator \eqref{alt2}-\eqref{alt1} is always applicable.
\end{remark}

Let us now provide a characterization of $DMSE_{N \infty}^{avg}(k)$ (and $DMSE^{avg}_{R,N \infty}(k)$), which of course follows immediately from our earlier characterization  of $MSE_{N \infty}^{avg}(k)$ (and $MSE^{avg}_{R,N \infty}(k)$). Immediately,
\begin{corollary} \textit{Under the conditions of Theorem \ref{bigtheorem}, we have that
		\begin{equation*}
				DMSE^{avg}_{N \infty}(k)=\frac{\overline{\sigma^2}(1-\overline{\rho})}{k(k+1)},
\end{equation*}
so that
$$DMSE^{avg}_{R,N \infty}(k)=\frac{DMSE^{avg}(k)}{DMSE^{avg}(1)}=\frac{2}{k(k+1)}.$$
In particular, because $DMSE^{avg}_{R,N \infty}(k)$ depends only on $k$ and not on the data, it will be identical for all variables forecast.
}
\end{corollary}

%
%
%
%
%
%
\begin{remark}
We earlier observed that the empirical $DMSE^{avg}_{R,N \infty}(k)$ signature plots for growth and inflation appeared identical. The corollary shows this is not a data-specific coincidence. Theoretically, the two curves \textit{must} coincide for every $k$.
\end{remark}

\begin{remark} Similarly, for both growth and inflation we earlier observed negligible diversification gains from expanding the forecast pool beyond $k=10$ (say), and the corollary shows why. By the time we reach $k=10$, the marginal gain from expanding the pool drops to just 1.8\% of the initial marginal gain $({2}/[{10 (10+1)]} = 0.018 )$. 
\end{remark}

\subsection{On the Assessment of Equicorrelation Model Fit}

As we have seen, equicorrelation ultimately emerges as a device for convenient calculation and understanding of direct signature plots, rather than as a separate ``model" producing separate signature plots. In particular, nothing we have done requires that the equicorrelation be ``true." Nevertheless, it may be of interest in some contexts to assess whether forecast errors are truly equicorrelated, which can be done (under stronger assumptions than those invoked thus far) by maximum-likelihood estimation of a dynamic-factor model, followed by likelihood-ratio tests of the equicorrelation restrictions.

Consider in particular a standard  dynamic single-factor model (DFM),
\vspace{-2mm}
\begin{equation} \label{m1}
	e_{it} = \delta_i z_{t} + w_{it} 
\end{equation}
\vspace{-10mm}
\begin{equation*}    
	z_{t} = \phi z_{t-1} + v_{t},
\end{equation*}
where $ w_{it} \sim iid \, N(0, \sigma_{wi}^{2})$, $v_{t} \sim iid \, N(0, \sigma_{v}^{2})$, and $w_{it} \perp v_{t}$,  $\forall i, t$.  The implied forecast-error covariance matrix $\Sigma$  fails to satisfy equicorrelation; that is,
\begin{equation*}  
	\Sigma ~ \neq ~  \sigma^2
	\begin{pmatrix}
		1 & \rho & \cdots & \rho \\
		\rho & 1 & \cdots & \rho \\
		\vdots & \vdots & \ddots & \vdots \\
		\rho & \rho & \cdots & 1
	\end{pmatrix},
\end{equation*}
because the forecast-error variances generally vary with $i$, and their correlations generally vary with $i$ and $j$. In particular, simple calculations reveal that
\begin{equation*} \label{sig1}
	\begin{split}
		\sigma^2_i \equiv var(e_{i,t}) &=   \delta_{i}^2 var(z_{t}) + \sigma_{wi}^{2} \\
		&=\delta_{i}^{2} \left( var(z_{t}) + \frac{\sigma_{wi}^{2}}{\delta_{i}^{2}} \right), ~ \forall i,
	\end{split}
\end{equation*} 
where $var(z_t) = \frac{\sigma_v^2}{1 - \phi^2}$, and
\begin{equation}
	\begin{split} \label{rho1}
		\rho_{ij} \equiv corr(e_{i,t}, e_{j,t})   &=  \frac{\delta_{i} \delta_{j} var(z_{t})}{\sqrt{\delta_{i}^{2} var(z_{t}) + \sigma_{wi}^{2}} \sqrt{\delta_{j}^{2} var(z_{t}) + \sigma_{wj}^{2}}} \\
		&= \frac{1}{\sqrt{1 + \frac{\sigma_{wi}^{2}}{\delta_{i}^{2}var(z_{t})}} \sqrt{1 + \frac{\sigma_{wj}^{2}}{\delta_{j}^{2}var(z_{t})}}} \,, ~ \forall i,j .
	\end{split}
\end{equation}

Nevertheless, certain simple restrictions on the DFM \eqref{m1} produce certain forms of equicorrelation. First, it is apparent from equation \eqref{rho1} that $\rho_{ij} = \rho,~ \forall i \ne j$, if and only if
\begin{equation}  \label{h0aa}
	\frac{\sigma_{wi}^{2}}{\delta_{i}^{2}} = \frac{\sigma_{wj}^{2}}{\delta_{j}^{2}}, ~ \forall i \ne j,
\end{equation}
so that imposition of the constraint \eqref{h0aa} on equation \eqref{m1} produces a ``weak" form of equicorrelation with identical correlations ($\rho$) but allowing for potentially different idiosyncratic shock variances ($\sigma_1^2, ..., \sigma_N^2$). That is, 
\begin{equation*} \label{equi2}
	\Sigma = 
	\begin{pmatrix}
		\sigma_1^2 & \rho & \cdots & \rho \\
		\rho & \sigma_2^2 & \cdots & \rho \\
		\vdots & \vdots & \ddots & \vdots \\
		\rho & \rho & \cdots & \sigma_N^2 \\
	\end{pmatrix}.
\end{equation*}

Second, it is also apparent from equation \eqref{rho1} that if we impose the stronger restriction, 
\begin{equation}         \label{strong33}
	\sigma_{wi}^{2} = \sigma_{wj}^{2} ~~ {\rm and} ~~ \delta_{i} = \delta_{j}, ~ \forall i, j,
\end{equation} 
which of course implies the weaker restriction \eqref{h0aa}, then we obtain (``strong") equicorrelation as we have defined it throughout this paper,  with identical correlations ($\rho$) \textit{and} identical idiosyncratic shock variances ($\sigma^2$). That is,
\begin{equation*} \label{equi3}
	\Sigma ~= ~  \sigma^2
	\begin{pmatrix}
		1 & \rho & \cdots & \rho \\
		\rho & 1 & \cdots & \rho \\
		\vdots & \vdots & \ddots & \vdots \\
		\rho & \rho & \cdots & 1
	\end{pmatrix}.
\end{equation*}

The DFM \eqref{m1} is already in state-space form, and one pass of the Kalman filter yields the innovations needed for exact Gaussian likelihood evaluation and optimization, and it also accounts for missing observations associated with survey entry and exit.  One may also impose the weak or strong  equicorrelation  restrictions \eqref{h0aa} or \eqref{strong33}, respectively, and assess them using likelihood-ratio tests. Unsurprisingly, such formal tests (not reported) reject equicorrelation -- a stylized ``toy" data-generating process if ever there was one -- for both growth and inflation, so we will not pursue formal testing.

\begin{figure}[tp]
	\caption{Growth and Inflation Forecast Error Variances and Covariances, Percent Deviations from Median}
	\label{assess}
	\centering
	\large
	
	Growth
	
	\includegraphics[trim={0mm 0mm 0 0},clip, scale=.6]{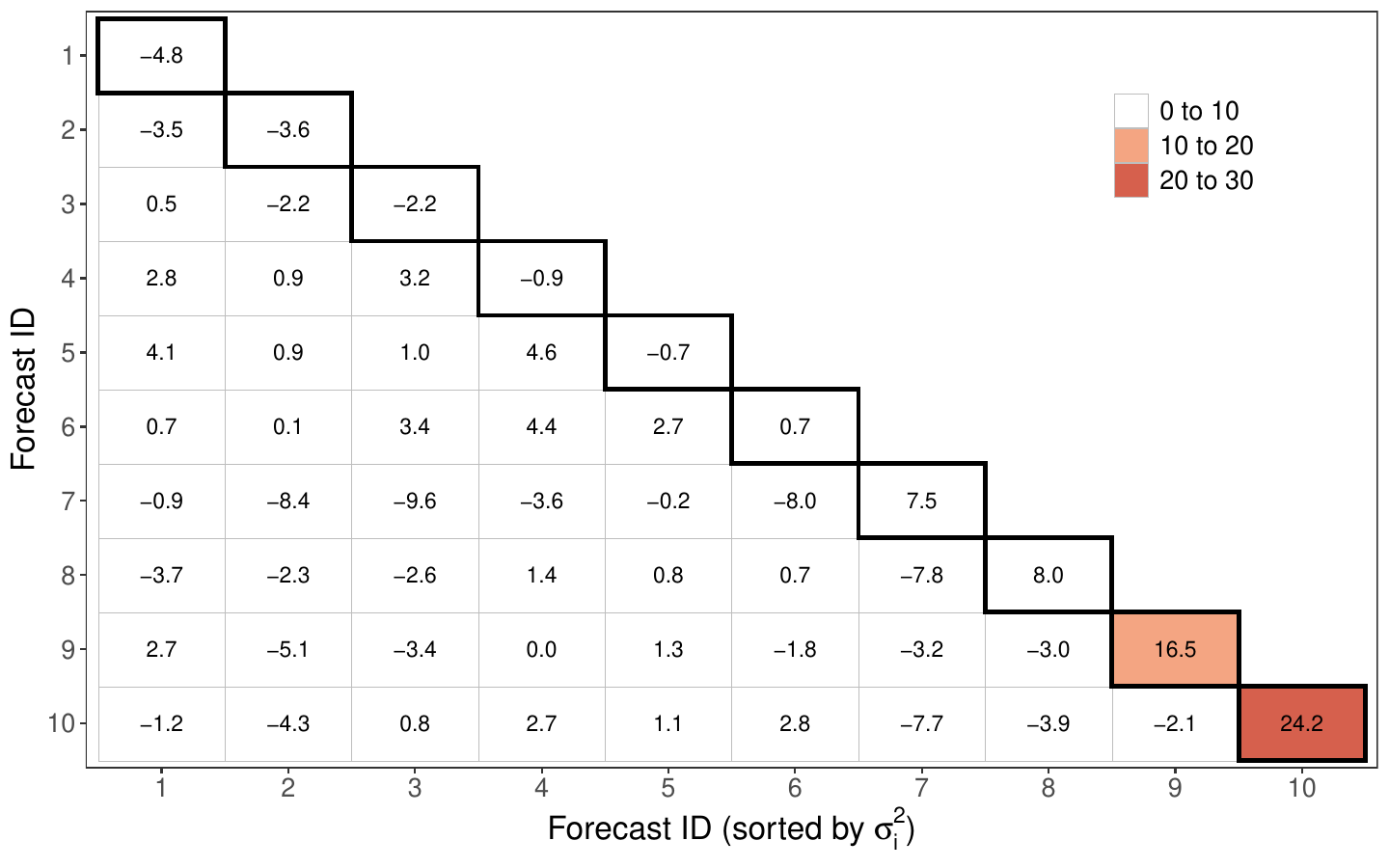}
	
	\vspace{6mm}
	
	Inflation
	
	\includegraphics[trim={0mm 0mm 0 0},clip, scale=.6]{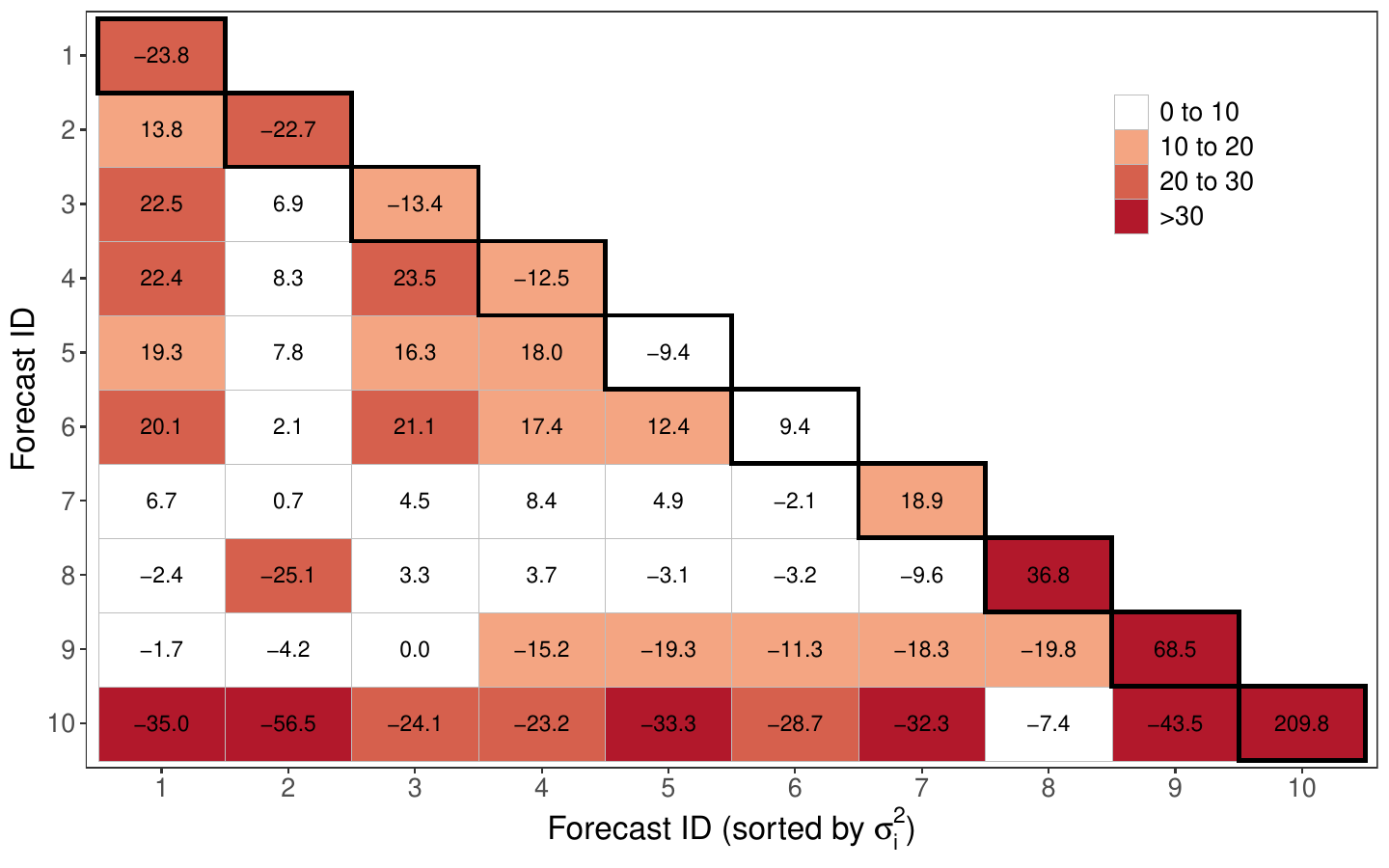}
	
	\begin{adjustwidth}{1cm}{1cm}
		\begin{spacing}{1.0}  \noindent  \footnotesize 
			Notes: We work with a balanced panel of ten forecasters, 2010Q1-2019Q4 (40 quarters). See text for details.
		\end{spacing}
	\end{adjustwidth}        
	
\end{figure}

More interestingly, we will provide a preliminary assessment of the \textit{size} of  deviations from equicorrelation, for both growth and inflation, quite apart from assessing whether they happen to be exactly zero. We construct a balanced panel with no entry or exit, 2010Q1-2019Q4 (40 quarters), containing ten forecasts for growth and inflation, and we examine their covariance matrices, as shown in Figure \ref{assess}.\footnote{The forecasts are ordered from lowest to highest error variance for each variable.}

First consider growth forecast errors, in the upper panel of Figure \ref{assess}. No shading means that the absolute deviation of the object in the cell (a variance or covariance) from the median across all cells (all variances or all covariances) is less than 10\%.  For example, the forecast with ID 3 has a variance of growth forecast error that is -2.2\% below the median variance, so its cell is unshaded. Similarly, light red shading means that the absolute deviation from the median is 10-20\%, and darker red shading means that the absolute deviation from the median is 20-30\%. All growth covariance cells are white, as are all but two variance cells. Hence, at least by our intentionally rough metric, equicorrelation appears to be not too bad an approximation for growth forecast errors.

Now consider inflation forecast errors, in the lower panel of Figure \ref{assess}. The shading convention is the same as earlier, but now there is a fourth, dark red, category corresponding to ${>}30\%$. The situation is sharply different, with many red cells, for both variances and covariances. This echoes our earlier discussion of the relative lack of consensus in inflation forecasts due to the difficulty of determining an appropriate inflation model, as per Figures \ref{fig: RMSE} and \ref{fig: boxplots}, together with Remarks \ref{remf} and \ref{rem}.


\section{Summary, Conclusions, and Directions for Future Research}  \label{concl}

We have  studied  the properties of macroeconomic survey forecast response averages as the number of survey respondents grows, characterizing the speed and pattern of the ``gains from diversification" and their eventual decrease with ``portfolio size" (the number of survey respondents) in both (1) the key real-world data-based environment of the U.S. Survey of Professional Forecasters (SPF), and (2) the theoretical model-based environment of equicorrelated forecast errors. We proceeded by proposing and comparing various  direct and model-based ``crowd size signature plots," which summarize the forecasting performance of $k$-average forecasts as a function of $k$, where $k$ is the number of forecasts in the average. We then estimated the equicorrelation model for growth and inflation forecast errors by choosing model parameters to minimize the divergence between direct and model-based signature plots. 

The results indicate near-perfect equicorrelation model fit for both growth and inflation, which we explicated by showing analytically that, under conditions, the direct and fitted equicorrelation model-based signature plots are identical at a particular model parameter configuration, which we characterize. We find that the gains from diversification are greater for inflation forecasts than for growth forecasts, but that both the inflation and growth diversification gains nevertheless decrease quite quickly (which we also explain analytically), so that fewer SPF respondents than currently used may be adequate.

Several directions for future research appear promising, including, in no particular order:

\begin{enumerate}
	\item Instead of considering $MSE$s across $N \choose k$ possible $k$-average forecasts and averaging to obtain a ``representative $k$-average" forecast $MSE$ as a function of $k$, one may want to consider ``\textit{best} $k$-average" forecast $MSE$ as a function of $k$, where the unique best $k$-average forecast is obtained in each period as the $k$-average that performed best historically. 

\item One may want to allow for time-varying equicorrelation parameters, as $\sigma^2$ might, for example, move downward with  the Great Moderation,  while $\rho$ might move counter-cyclically. The strong equicorrelation model in dynamic-factor form becomes 
\vspace{-2mm}
\begin{equation*} \label{m3}
        e_{it} = \delta_t z_{t} + w_{it} 
\end{equation*}
\vspace{-10mm}
\begin{equation*}  \label{m4}
        z_{t} = \phi z_{t-1} + v_{t},
\end{equation*}
where $ w_{it} \sim iid (0, \sigma_{wt}^{2})$, $v_{t} \sim iid (0, \sigma_{v}^{2})$, and $w_{it} \perp v_{t},  \forall i, t$. Immediately, 
$$
e_t \sim iid \left (0,~ \Sigma_t (\rho_t) \right ),
$$  
where
\begin{equation*} 
        \Sigma_t(\rho_t) ~=~  \sigma_t^2
        \begin{pmatrix}
                1 & \rho_t & \cdots & \rho_t \\
                \rho_t & 1 & \cdots & \rho_t \\
                \vdots & \vdots & \ddots & \vdots \\
                \rho_t & \rho_t & \cdots & 1
        \end{pmatrix}
\end{equation*}
$$
\sigma_t^2 = \delta_t^2 var (z_t) + \sigma_{wt}^2
$$
$$
\rho_{t} = \frac{\delta_t^2 var (z_t)}{\delta_t^2 var (z_t) +  \sigma_{wt}^2}.
$$

\item One may want to complement our exploration of the U.S. SPF with a comparative exploration of the European SPF.\footnote{For an introduction to the European SPF, see the materials at  \url{https://data.ecb.europa.eu/methodology/survey-professional-forecasters-spf}.} Doing so appears feasible but non-trivial, due to cross-survey differences in sample periods, economic indicator concepts (e.g., inflation), and timing conventions.  

\end{enumerate}

 
\appendix
\appendixpage
\addappheadtotoc

\newcounter{saveeqn}
\setcounter{saveeqn}{\value{section}}
\renewcommand{\theequation}{\mbox{\Alph{saveeqn}\arabic{equation}}} \setcounter{saveeqn}{1}
\setcounter{equation}{0}

\allowdisplaybreaks

\section{Data Definitions and Sources}
\label{app_data}

\setcounter{saveeqn}{\value{section}}
\renewcommand{\theequation}{\mbox{\Alph{saveeqn}\arabic{equation}}} \setcounter{saveeqn}{1}
\setcounter{equation}{0}

We obtain U.S. quarterly level forecasts of real output and the GDP deflator from the Federal Reserve Bank of Philadelphia's \textit{Individual Forecasts: Survey of Professional Forecasters} (variables $RGDP$ and $PGDP$, respectively).  We  transform the level forecasts into annualized growth rate forecasts using:
\begin{equation} \label{ann}
g_{t+h|t-1}=100\left(\left(\frac{f_{t+h|t-1}}{f_{t+h-1|t-1}}\right)^{4}-1\right),
\end{equation}
where $f_{t+h|t-1}$ is a quarterly level forecast (either $RGDP$ or $PGDP$) for quarter $t+h$ made using  information available in quarter $t-1$. For additional information, see \url{https://www.philadelphiafed.org/surveys-and-data/real-time-data-research/individual-forecasts}.

We obtain the corresponding realizations from the Federal Reserve Bank of Philadelphia's   \textit{Forecast Error Statistics for the Survey of Professional Forecasters} (December 2023 vintage).  The  realizations are reported as annualized growth rates, as in equation \eqref{ann} above, so there is no need for additional transformation.  For additional information, see \url{https://www.philadelphiafed.org/surveys-and-data/real-time-data-research/error-statistics}.

\section{Optimal Combining Weights Under Weak Equicorrelation} \label{genequi}

\setcounter{saveeqn}{\value{section}}
\renewcommand{\theequation}{C\arabic{equation}} \setcounter{saveeqn}{1}
\setcounter{equation}{0}

Here we briefly consider the ``weak equicorrelation" case, with  correlation $\rho$ and variances $\sigma_1^2,~ \sigma_{2}^{2},~ ..., ~ \sigma_{N}^{2}  $; that is,
\[
\Sigma = 
\begin{pmatrix}
\sigma_{1}^{2} & \rho & \cdots & \rho \\
\rho & \sigma_{2}^{2} & \cdots & \rho \\
\vdots & \vdots & \ddots & \vdots \\
\rho & \rho & \cdots & \sigma_{N}^{2} \\
\end{pmatrix},
\]
where $\rho \in \left ] \frac{-1}{N-1}, 1 \right [$. We can decompose the covariance matrix $\Sigma$ as
\[
\Sigma = DRD = 
\begin{pmatrix}
\sigma_{1} & 0 & \cdots & 0 \\
0 & \sigma_{2} & \cdots & 0 \\
\vdots & \vdots & \ddots & \vdots \\
0 & 0 & \cdots & \sigma_{N} \\
\end{pmatrix}
\begin{pmatrix}
1 & \rho & \cdots & \rho \\
\rho & 1 & \cdots & \rho \\
\vdots & \vdots & \ddots & \vdots \\
\rho & \rho & \cdots & 1 \\
\end{pmatrix}
\begin{pmatrix}
\sigma_{1} & 0 & \cdots & 0 \\
0 & \sigma_{2} & \cdots & 0 \\
\vdots & \vdots & \ddots & \vdots \\
0 & 0 & \cdots & \sigma_{N} \\
\end{pmatrix},
\]
 where $R$ is positive definite if and only if $\rho \in \left] \frac{-1}{N-1},1 \right [ $.  The inverse of the covariance matrix is 
\[
\Sigma^{-1} = D^{-1}R^{-1}D^{-1} = 
\begin{pmatrix}
\sigma_{1}^{-1} & 0 & \cdots & 0 \\
0 & \sigma_{2}^{-1} & \cdots & 0 \\
\vdots & \vdots & \ddots & \vdots \\
0 & 0 & \cdots & \sigma_{N}^{-1} \\
\end{pmatrix}
\begin{pmatrix}
1 & \rho & \cdots & \rho \\
\rho & 1 & \cdots & \rho \\
\vdots & \vdots & \ddots & \vdots \\
\rho & \rho & \cdots & 1 \\
\end{pmatrix}^{-1}
\begin{pmatrix}
\sigma_{1}^{-1} & 0 & \cdots & 0 \\
0 & \sigma_{2}^{-1} & \cdots & 0 \\
\vdots & \vdots & \ddots & \vdots \\
0 & 0 & \cdots & \sigma_{N}^{-1} \\
\end{pmatrix},
\] 
where
\[
R^{-1} = \frac{1}{1-\rho} I - \frac{\rho}{(1-\rho)(1+(N-1)\rho)} \iota \iota',
\]
 $I$ stands for an $N\times N$ identity matrix, and $\iota$ is a $N$-vector of ones.

 Recall that, as noted in the text, the optimal combining weight is
  \begin{equation}  \label{optimal1app}
        \lambda^* = \left (\iota' \Sigma^{-1} \iota \right)^{-1}  \Sigma^{-1} \iota,
 \end{equation}
 The first part  of the optimal combining weight  \eqref{optimal1app} is
\begin{equation} \label{eq: optimal weights 1}
\iota' \Sigma^{-1} \iota  = \frac{(1+(N-1)\rho) \left(\sum_{i=1}^{N} \sigma_{i}^{-2}\right) - \rho \left(\sum_{i=1}^{N} \sigma_{i}^{-1} \right)\left(\sum_{i=1}^{N} \sigma_{i}^{-1} \right)}{(1-\rho)(1+(N-1)\rho)},
\end{equation}
and the second part is
\begin{equation} \label{eq: optimal weights 2}
\Sigma^{-1} \iota = \frac{(1+(N-1)\rho) \boldsymbol{\sigma}^{-2} - \rho \left(\sum_{i=1}^{N} \sigma_{i}^{-1}\right) \boldsymbol{\sigma}^{-1}}{(1-\rho)(1+(N-1) \rho)},
\end{equation}
where 
\[
\boldsymbol{\sigma}^{-2} = 
\begin{pmatrix}
\sigma_{1}^{-2} \\
\sigma_{2}^{-2} \\
\vdots \\\
\sigma_{N}^{-2}
\end{pmatrix}
\quad \text{and} \quad
\boldsymbol{\sigma}^{-1} = 
\begin{pmatrix}
\sigma_{1}^{-1} \\
\sigma_{2}^{-1} \\
\vdots \\\
\sigma_{N}^{-1}
\end{pmatrix}.
\]
Inserting equations \eqref{eq: optimal weights 1} and  \eqref{eq: optimal weights 2} into equation \eqref{optimal1app}, we get the optimal weight for the $i$th forecast as
\begin{equation} \label{eq: optimal weights 3}
\lambda_{i}^{*} = \frac{\sigma_{i}^{-2} +  \rho(N-2) \sigma_{i}^{-2} - \rho \left( \sum_{j\neq i} \sigma_{i}^{-1} \sigma_{j}^{-1} \right)}{\sum_{i=1}^{N} \left(\sigma_{i}^{-2} +  \rho(N-2) \sigma_{i}^{-2} - \rho \left( \sum_{j\neq i} \sigma_{i}^{-1} \sigma_{j}^{-1} \right) \right)}.
\end{equation}
To check the formula, note that for $N=2$ we obtain the standard \cite{BatesandGranger1969} optimal bivariate combining weight,
\[
\lambda_{1}^{*} = \frac{\sigma_{1}^{-2} - \rho \sigma_{1}^{-1} \sigma_{2}^{-1}}{ \sigma_{1}^{-2} + \sigma_{2}^{-2} - 2\rho \sigma_{1}^{-1} \sigma_{2}^{-1} } = \frac{\sigma_{2}^{2} - \rho \sigma_{1} \sigma_{2}}{\sigma_{1}^{2} + \sigma_{2}^{2} - 2 \rho \sigma_{1} \sigma_{2}},
\]
and for any $N$, but with  $\sigma^2_{j} = \sigma^2$  $\, \forall j$ (equicorrelation), we  obtain weights,
\[
\lambda_{i}^{*} = \frac{ (1-\rho) \sigma^{-2}}{N (1-\rho) \sigma^{-2}} = \frac{1}{N}, ~~ \forall i.
\]

\addcontentsline{toc}{section}{References}
\bibliographystyle{Diebold}
\bibliography{Diebold}

\end{document}